# Sliding-mediated ferroelectric phase transition in $CuInP_2S_6$ under pressure

Zhou Zhou[1,#], Jun-Jie Zhang[2,#], Gemma F. Turner[3,#], Stephen A. Moggach[3,*], Yulia Lekina[4], Samuel Morris[5], Shun Wang[1], Yiqi Hu[1], Qiankun Li[1], Jinshuo Xue[1], Zhijian Feng[1], Qingyu Yan[1], Yuyan Weng[1], Bin Xu[1], Yong Fang[6], Ze Xiang Shen[4], Liang Fang[1,*], Shuai Dong[2,*], Lu You[1,*]

[1]School of Physical Science and Technology, Jiangsu Key Laboratory of Thin Films, Soochow University, Suzhou, 215006, China

[2]Key Laboratory of Quantum Materials and Devices of Ministry of Education, School of Physics, Southeast University, Nanjing 211189, China

[3]School of Molecular Sciences, The University of Western Australia, 35 Stirling Highway, Crawley, Perth, Western Australia 6009, Australia.

[4]Centre for Disruptive Photonic Technologies, Nanyang Technological University, 637371, Singapore

[5]Facility for Analysis, Characterisation, Testing and Simulation (FACTS), Nanyang Technological University, 639798, Singapore

[6]Jiangsu Laboratory of Advanced Functional Materials, School of Electronic and Information Engineering, Changshu Institute of Technology, Changshu, 215500, China

#These authors contributed equally.

*Corresponding authors:

stephen.moggach@uwa.edu.au

lfang@suda.edu.cn

sdong@seu.edu.cn

lyou@suda.edu.cn

## Abstract

Interlayer stacking order has recently emerged as a unique degree of freedom to control crystal symmetry and physical properties in two-dimensional van der Waals (vdW) materials and heterostructures. By tuning the layer stacking pattern, symmetry-breaking and electric polarization can be created in otherwise non-polar crystals, whose polarization reversal depends on the interlayer sliding motion. Herein, we demonstrate that in a vdW layered ferroelectric, its existing polarization is closely coupled to the interlayer sliding driven by hydrostatic pressure. Through combined structural, electrical, vibrational characterizations, and theoretical calculations, we clearly map out the structural evolution of $CuInP_2S_6$ under pressure. A tendency towards a high polarization state is observed in the low-pressure region, followed by an interlayer-sliding-mediated phase transition from a monoclinic to a trigonal phase. Along the transformation pathway, the displacive-instable Cu ion serves as a pivot point that regulates the interlayer interaction in response to external pressure. The rich phase diagram of $CuInP_2S_6$, which is enabled by stacking orders, sheds light on the physics of vdW ferroelectricity and opens an alternative route to tailoring long-range order in vdW layered crystals.

## Introduction

Recent developments in two-dimensional van der Waals (vdW) ferroelectric materials have brought unexpected twists to ferroelectric physics[1,2]. The reduced lattice dimensionality and large degrees of freedom for layer stacking enable unconventional properties in two-dimensional (2D) ferroelectrics in comparison to their three-dimensional (3D) counterparts, such as sliding/moiré ferroelectricity[3-6], layer-dependent cumulative polarization[7,8]. As a classic vdW layered crystal with robust room-temperature ferroelectricity, $CuInP_2S_6$ (CIPS) has received revived interest recently, due to its fascinating ferroelectric characteristics[9,10]. Most notable examples include negative longitudinal piezoelectricity[11,12], strongly intertwined polarization switching and ionic conductivity[13,14], as well as quadruple-well polarization states[15]. All these unique features stem from the displacive instability of the monovalent $Cu^+$ in the 2D crystal lattice. Hence, thorough investigations of the displacement behavior of $Cu^+$ in CIPS under electric, strain, and temperature stimuli hold great importance for understanding the ferroelectric physics in this intriguing compound, and may deliver more general insights into relevant 2D ferroelectrics.

One particular point of interest, is the metastable Cu site within the vdW gap, which lies at the core of the observed negative $d_{33}$ and strain-tunable quadruple-well ferroelectricity. Previous work demonstrated that it is possible to stabilize the occupancy of the in-gap Cu site by heterostructure strain in a mixed-phase $CuInP_2S_6$-$In_{4/3}P_2S_6$ solid solution, resulting in a high-polarization state with inverse piezoelectric sign[15]. However, the inhomogeneous composition and strain distribution across the sample complicates the intrinsic property characterizations, and one may naturally beg the question: can we realize the high polarization state by applying a homogeneous hydrostatic stress? To this end, it is necessary to study how phase pure CIPS behaves under high pressure. Unlike conventional positive-$d_{33}$ ferroelectrics, whose polarization decreases and thus ferroelectric order deteriorates on the application of pressure[16], CIPS should behave oppositely[17]. In fact, early high-pressure work on CIPS discovered a first-order phase transition at ~4.0 GPa[18], with a potential symmetry change from monoclinic to trigonal. Later, the dependence of dielectric properties of CIPS on application of pressure was studied[19,20], which revealed an enhancement of the Curie temperature ($T_C$) as expected for a ferroelectric with negative piezoelectricity. However, detailed structural and polarization evolutions of CIPS under hydrostatic pressure remains unclear. Herein we have studied the effect of pressure on CIPS using single-crystal high-pressure X-ray diffraction (XRD), electrical conductivity, Raman spectroscopy and theoretical calculations, in order to map out the structural evolution of $CuInP_2S_6$ and its subsequent physical properties.

## Results

### Structural evolution of CIPS under hydrostatic pressure

CIPS crystallizes in a vdW monolayered structure, the structural backbone of which is an edge-shared $S_6$ octahedral framework, inside which Cu, In ions and P-P pairs fill in order, forming a triangular lattice. The second layer stacks atop by first rotating 180° and shifting by $1/3\vec{a}$ along the ***a***-axis, resulting in monoclinic lattice symmetry (space group *Cc*) with $\vec{b}$ as the unique axis[21]. Displacement of the Cu atom from the octahedral center is the main source of ferroelectricity for CIPS. With increasing temperature, the Cu ion exhibits strong thermal broadening perpendicular to the basal plane[22]. Both experimental and theoretical studies have shown that this thermal hopping effect is accompanied with an additional metastable site within the vdW gap[12,15], analogous to the second-order Jahn Teller effect found in zinc-blende CuCl[23]. Therefore, the occupancy of the Cu is presented as two independent sites (**Figure 1a**), of which one is located within the $S_6$ octahedron, and referred to as 'inside Cu' ($Cu^I$), whereas the other Cu-site, sits within the vdW gap ('outside Cu' ($Cu^O$)). In other words, the $Cu^O$ site is stabilized by the interlayer interaction with the S atom across the vdW gap.

The structural evolution of CIPS on increasing pressures at room temperature is shown in **Figure 1a**, as obtained from single-crystal XRD measurements. At the early stage of the pressurization, the basic crystal structure of CIPS remains unchanged. However, the relative population of the Cu ions between the two sites varies rapidly on increasing pressure. Particularly, the fraction of the $Cu^O$ site increases continuously, from 10 % under ambient pressure to ≈ 60 % at ~1 GPa (**Figure 1e**). On increasing pressure further, the Cu occupancies stay insensitive to the pressure change. At ~4 GPa, the pressure-induced structural phase transition occurs. The adjacent layers slide by $1/3\,\vec{a}$ along the ***a***-axis, transforming from monoclinic to trigonal symmetry (space group $P\bar{3}1c$). Due to the interlayer sliding, the Cu column is no longer aligned with the S4 atom of the adjacent layer (labelled S4 in **Figure 1a**), but in line with the P-P column. Consequently, the stacking sequence of the sulphur layers switches from ABC-type to AB-type closed pack (**Figure S1**). Meanwhile, the Cu ions jump back into the sulphur octahedral cage, with the majority staying at the central site and the rest distributing symmetrically between the upper and lower $Cu^I$ sites. As a result, CIPS now becomes a centrosymmetric non-polar phase.

On increasing pressure to ≈ 3.5 GPa, and prior to the phase transition, CIPS compresses hydrostatically, with a gradual reduction in cell volume (**Figure 1c**). The compressibility within the plane of the intralayer is smaller than that of the ***c**** axis (normal to the plane), whose length is reduced by 5 % up to ~ 3.4 GPa (**Figure 1b**). This is consistent with the highly anisotropic nature of chemical connectivity, as a result of the vdW layered structure. At the phase transition, there is a sudden contraction of ≈ 2 % along the ***c**** direction which is accompanied with interlayer sliding. The interlayer bond between Cu and S4 is also broken due to the sliding (average bond length changes from ≈ 2.63 to ≈ 3.29 Å), resulting in further collapse of the vdW gap thickness (**Figure 1e**). Concomitantly, the cell volume also exhibits a discontinuous jump at the phase transition boundary (**Figure 1c**). The bulk modulus $K_0$ is estimated to be 25.2(3.3) GPa, consistent with previous report[24]. The monoclinic angle $\beta$ also increases slightly on increasing pressure (**Figure 1d**), as a result of the weak sliding between adjacent layers. By scrutinizing the evolutions of layer and vdW gap thicknesses (defined by the average z position of the sulphur planes), how the structure of CIPS responds to compression can be followed (**Figure 1e**). The initial drop of layer thickness and rise of the gap thickness are in accordance with the drastic relocation of $Cu^{I}$ towards $Cu^{O}$ sites at lower pressures (**Figure 1f**). Further pressurization barely affects the layer thickness, but compresses the gap between the layers. On undergoing the phase transition, there is a sudden reduction in the gap thickness, consistent with the change in the lattice parameter along ***c***-axis. On the contrary, the layer thickness expands in the trigonal phase, because all the Cu ions now jump back into the octahedral cage. As an indicator of the polarization magnitude, the average off-center displacement of Cu ion with respect to the octahedral center is also derived, which shows a similar trend as the pressure-dependent $Cu^{O}$ occupancy (**Figure 1g**). The enhanced off-center displacement of Cu ion also results in strong puckering of the sulphur plane, due to the contraction of the Cu-S4 bond (**Figure S2b**). Structural distortions of the sulphur octahedra also takes place, as manifested by the expansion and contraction of the upper and lower $S_3$ triangle in the $CuS_6$ sublattice as well as the counter-rotations of the In and P-P octahedra (**Figure S2e**). These structural distortions are visibly suppressed in the high-pressure trigonal phase (**Figure S2c & f**). Previous theoretical calculation also confirmed a small but nonzero in-plane polarization component ($P_x$) in CIPS along the ***a***-axis[25]. Structurally, $P_x$ is associated with the relative in-plane displacement ($x_{Cu\text{-}S4}$) between Cu and the S4 atom. Under pressurization, owing to the interlayer-sliding, $x_{Cu\text{-}S4}$ gradually decreases, suggesting the reduction of $P_x$ (**Figure S3**).

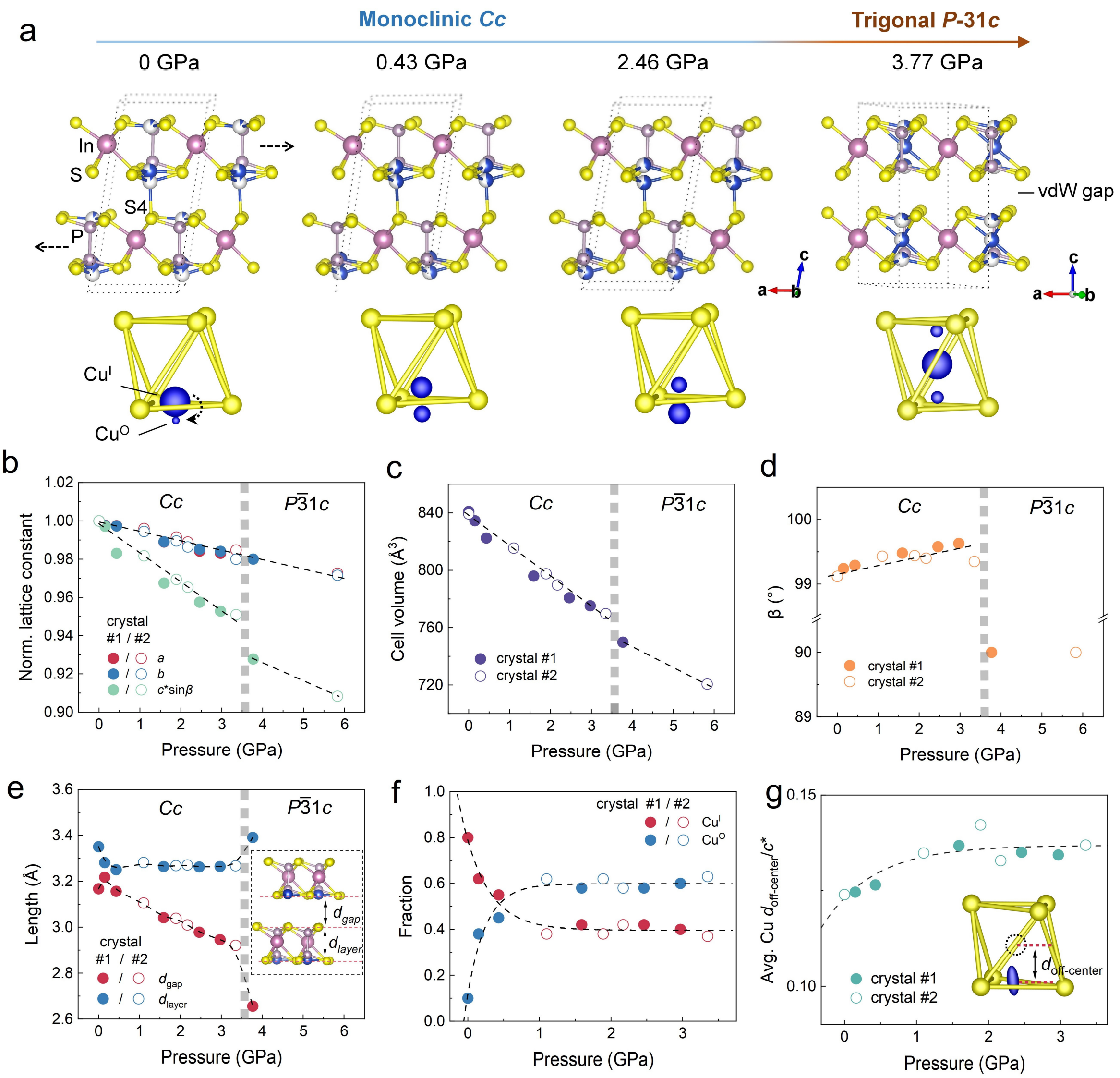

**Figure 1. Structural evolution of CIPS under hydrostatic pressure.** (a) Upper panel: the evolution of the crystal structure of CIPS with hydrostatic pressure. The pink, purple, and yellow balls represent In, P, and S atoms, respectively. The blue-white balls are Cu sites and the proportion of blue colors indicates the corresponding occupancy of the respective site. Lower panel: zoomed-in image of the $CuS_6$ sublattice. The ball size of the Cu represents the corresponding occupancy of the site. (b) The lattice parameters, (c) cell volume, (d) monoclinic angle $\beta$, (e) layer and gap thicknesses, (f) the fraction of $Cu^I$ and $Cu^O$ sites and (g) average off-center displacement of Cu ion as a function of the applied hydrostatic pressure.

To further understand above experimental discovery, we performed first-principles density functional theory (DFT) calculations. The possible phase transition caused by pressure in CIPS is determined through a comparison of enthalpies of Cc and $P\bar{3}1c$ phase. The enthalpy as function of pressure is calculated based on H = E + PV, where E is the total DFT energy, P is pressure and V is volume of unit cell. Our calculation indeed finds that CIPS undergoes a pressure-induced phase transition at ~5.8 GPa (at 0 K), from monoclinic (space group *Cc*) to trigonal phase (space group $P\bar{3}1c$) (**Figure S4c**). In the calculation, the $P\bar{3}1c$ phase is the high-symmetry paraelectric phase with Cu lying at the center of the $S_6$ octahedron. However, DFT results reveal that the $P\bar{3}1c$ phase exhibits one unstable phonon mode at Γ point, which is mainly contributed from Cu displacements, consequently leading to spontaneous phase transition from paraelectric $P\bar{3}1c$ to polar $P31c$ structure (**Figure S4b**). The calculated energy difference between the $P31c$ and $P\bar{3}1c$ phase is only 28.7 meV/f.u.. Thus, it is clear that the ground state (at 0 K) of CIPS under high pressure is a polar ferroelectric, while it turns to be the paraelectric one at room temperature as experimentally observed. A rough estimation of the ferroelectric phase transition temperature is 51 K (see Discussion for details). The critical pressure in our calculation is close but slightly higher than the experimental one, which is also reasonable considering the thermal effect in room temperature experiments.

Our DFT calculation also shows that the Cu atoms occupy the $Cu^{I}$ site below 0.5 GPa, before preferring the $Cu^{O}$ site above 1 GPa, a similar trend of Cu displacement (compared to the average Cu position) from the experimental data (**Figure S4d**). However, thermal hopping of the Cu ion causes a mixed occupancy between the $Cu^{I}$ and $Cu^{O}$ sites, leading to a broadening of its distribution and smaller displacement of the center of mass observed at room temperature.

**Electrical measurements under hydrostatic pressure**

To probe the evolution of the ferroelectric polarization under pressure, polarization-voltage hysteresis loops of the CIPS single crystal were measured during the process of compression (**Figure 2a**), from 0 to ≈ 1.8 GPa. The remanent hysteresis technique was used to remove the contribution from non-remanent polarization components and leakage due to ionic conduction. The polarization value increases with the increment of pressure, consistent with the negative piezoelectricity of CIPS. In the early stage of pressurization, the polarization value rises rapidly, from 4.4 μC/cm$^2$ at ambient pressure to about 6.0 μC/cm$^2$ under 0.5 GPa (**Figure 2b**). Afterwards, the growth rate slows down. The polarization value slowly increases at a rate of about 0.7 μC/cm$^2$ per GPa. This behavior is in good agreement with the trend of $Cu^{I}$ and $Cu^{O}$ site fraction obtained from experimental XRD refinements (**Figure 1f**), from which the average off-center displacement of the Cu ion can be observed (**Figure 1g**). The results indicate that the initial increase of the polarization is due to the sharp transfer of $Cu^{I}$ ions to $Cu^{O}$ site at the low-pressure range, whereas the slow increase of polarization at high pressure can be attributed to the volumetric effect. In this case, the relative displacement of Cu from the center of sulphur octahedron remains almost fixed (clamped-ion), but the volume of the lattice continues to shrink, leading to enhanced polarization (dipole moment per volume). Real and imaginary parts of relative dielectric permittivity are also measured as a function of pressure and plotted in temperature and frequency domains (**Figure S5a**). **Figure 2c** shows the temperature-dependent real part of the permittivity measured at a frequency of 950 kHz under a pressure of 0, 0.045, 0.09, 0.128 GPa, respectively. Within our measurement range, the Curie temperature ($T_C$) of CIPS rises linearly at a rate of about 353.6 K/GPa (**Figure 2d**). This pressure coefficient is much larger than previously reported[18,19], potentially due to the inaccuracy in the pressure calibration and highly hysteretic behavior in this low-pressure region. Since CIPS exhibits finite ionic conduction at room temperature, through the dielectric data, we can obtain the AC conductivity $\sigma = 2\pi f \varepsilon_0 \varepsilon_r''$ (**Figure S5b**), where $f$ is the measurement frequency, $\varepsilon_r''$ is the imaginary part of the relative permittivity of CIPS and $\varepsilon_0$ is the vacuum permittivity. The DC conductivity $\sigma_{dc}$ can be extracted by fitting the low-frequency part of the curve according to Jonscher's power law[26]:

$$\sigma = \sigma_{dc} + A(2\pi f)^s \,, \tag{1}$$

from which $A$ is the pre-exponential constant and $s$ is the power law exponent with $0 < s < 1$. Finally, its activation energy $E_A$ of the DC ionic conduction is given based on the Arrhenius equation[27]:

$$\sigma T = \sigma_0 \cdot e^{\frac{-E_a}{k_B T}} \,, \tag{2}$$

and an Arrhenius plot demonstrates different $E_A$ under different pressures (**Figure 2e**). The activation energy of CIPS increases with higher pressure below the Curie temperature. This means that compression inhibits ionic conductivity, which is also supported by the reduction of the leakage signal in the low-frequency polarization hysteresis loops under pressure (**Figure S5c**). The pressure-suppressed ionic conduction can be understood by an increased energy barrier for thermal hopping of the Cu ions due to a more compact crystal lattice, as also evidenced by the decrease of the thermal parameters $U_{iso}$ (Cu) derived from the structural refinements (**Figure 2f**). The increased sharpness of the polarization hysteresis loop under pressure can also be explained by the suppression of the Cu hopping. The final rise of the $U_{iso}$ (Cu) can be attributed to the proximity to critical pressure for the structural phase transition.

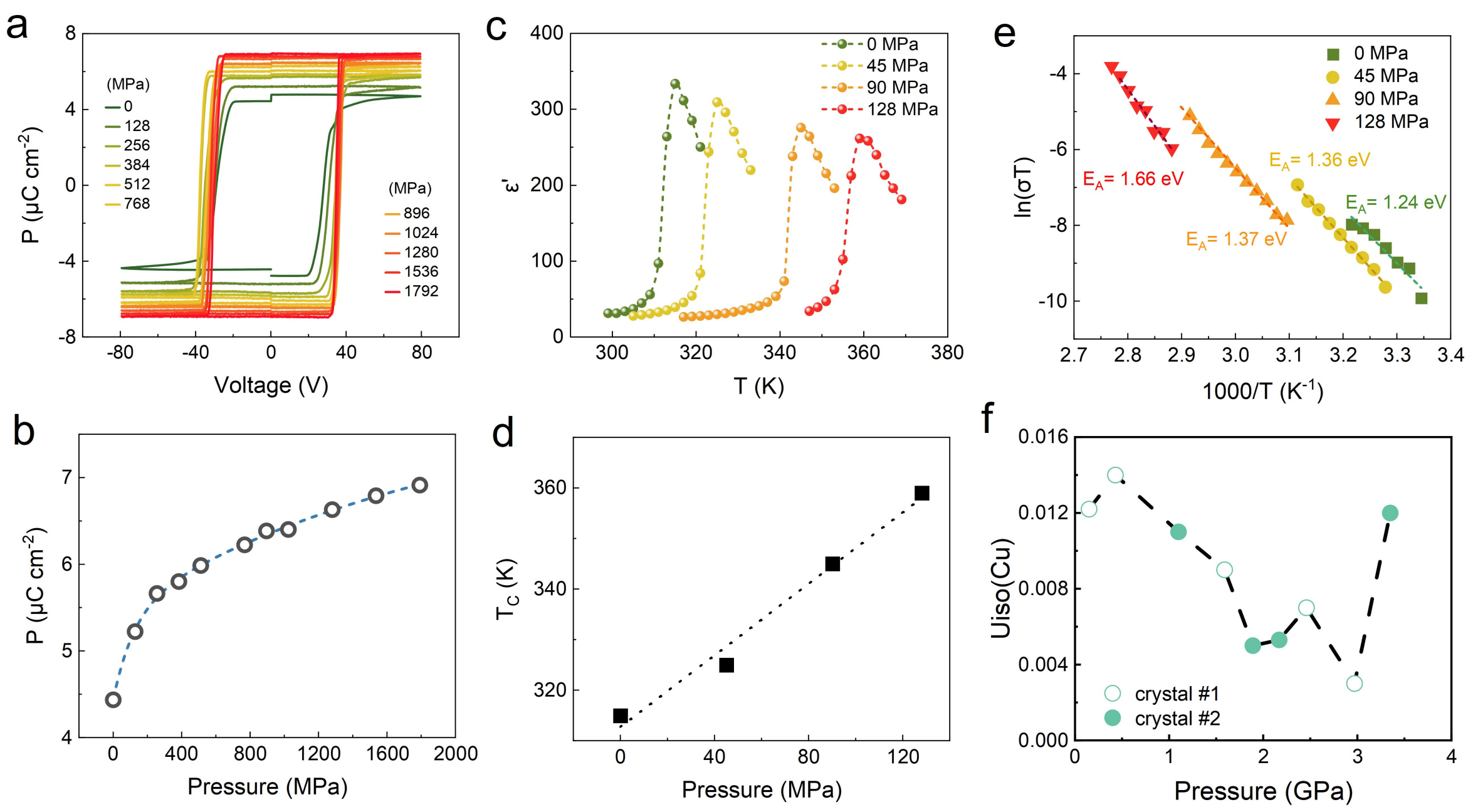

**Figure 2. Pressure-dependent electrical measurements.** (a) Remanent polarization hysteresis loops of CIPS crystal measured at different pressures. The measurements were carried out at room temperature with a triangular voltage waveform of 12.5 Hz. (b) Measured remanent polarization as a function of applied hydrostatic pressure. (c) The dielectric anomalies of real permittivity near the ferroelectric-paraelectric phase transition boundary measured under 0, 45, 90, 128 MPa with a frequency of 950 kHz. (d) The Curie temperature as a function of applied hydrostatic pressure. (e) The Arrhenius plot of the DC conductivity with extracted activation energies under 0, 45, 90, 128 MPa (below $T_C$). (f) Thermal parameters of Cu atoms under different pressures derived from the structural refinements.

## Vibrational spectrum of CIPS under high pressure

Raman spectroscopy was employed to probe kinetic changes of phonon modes of single crystal CIPS on increasing pressure. The primitive cell of CIPS contains two formula units with 20 atoms, which results in a total of 60 normal vibrational modes (including 3 acoustic modes). Based on the group-theoretical analysis, the irreducible representations for all optical vibrational modes at the Brillouin zone center (Γ) are[28]:

$$\Gamma_{optic} = 28A' + 29A'', \quad (3)$$

among which the non-degenerate A' (symmetric) and A'' (antisymmetric) are both Raman and infrared active. We have calculated all the phonon modes and their corresponding scattering intensity based on calculated Raman tensors at 0 K, as summarized in **Table S1**. The simulated Raman spectrum (light polarization along the [110] direction) generally matches the experimental result (light polarization within *ab* plane) except for some red-shifts of the high-frequency modes (**Figure S6a**). Notably, the low-frequency peak not predicted by the theory can be attributed to the out-of-plane component of atomic vibrations due to random light scattering at the crystal edges[24]. By adding some out-of-plane electric field component, the low-frequency peak can be nicely reproduced by the simulated spectrum (**Figure S6b**). The majority of the Raman active peaks appear to blueshift on increasing pressure due to an overall hardening of the elastic constant of the chemical bonds induced by lattice compression (**Figure 3a**). Generally, the peak shift behavior can be classified into two categories: one group (73, 103, 164, 266, 375 $cm^{-1}$ bands) shows a sudden rise after 0.2 GPa, whereas the other group (216, 239, 319 $cm^{-1}$ bands) softens first and hardens after 0.5 GPa (**Figure S7c-e**). The kink at 0.2 – 0.5 GPa can be attributed to the abrupt occupancy transfer from $Cu^{I}$ to $Cu^{O}$ site. The hardening of the former group can be understood by the enhanced interlayer interaction by more $Cu^{O}$ occupancy. Through a careful analysis of the atomic motions for different phonon modes (**Supplementary gif images**), it is found that the modes in the latter group all involve either the bending or stretching motion of the $Cu\text{-}S_3$ bonds. During the $Cu^{I}$ to $Cu^{O}$ transition, the force constant of the $Cu\text{-}S_3$ bonds is possibly reduced, thus leading to a softening behavior of related vibrational modes.

When the pressure is greater than ≈ 4 GPa, the number of Raman peaks decreases significantly, signaling the occurrence of the phase transition and higher symmetry of the high-pressure phase. This finding is consistent with the XRD results. In decompression branch, the phase transformation occurs at lower pressure, suggesting first-order phase transition with hysteresis (**Figure S7a, b**). It's worth noting that a softening trend of the low-frequency peak is found, in stark contrast to other peaks (**Figure 3b**). This peak can be further deconvoluted into two peaks by Gaussian multi-peak fitting (**Figure 3c and Figure S8**), and we found completely opposite peak shifts for these two peaks with regard to pressure, namely, one hardens, while the other softens (**Figure 3d**).

Our DFT calculations reveal that the main contribution of Raman peak at low frequency comes from mode #7 (**Figure 3f**), that is, the opposite vibration of adjacent layers in the z direction, including the z direction vibration of Cu atoms and a slight interlayer sliding. The pressure-dependent behavior of this vibration mode appears to directly correlate with the Cu position within the lattice (**Figure 3g**). Below 1 GPa, the Cu remains inside the sulphur octahedron ($Cu^I$ site). Meanwhile, mode #7 continues to soften in this pressure range. Above 1 GPa, the Cu ion suddenly jumps outside the octahedral cage and into the vdW gap ($Cu^O$ site). Correspondingly, mode #7 starts to harden with increasing pressure. Based on these calculation results, we can conclude that the two sub-peaks around 30 $cm^{-1}$ correspond to the contribution of $Cu^I$ and $Cu^O$ site occupancies, respectively. The opposite pressure-dependent behaviors for these two sites result in the crossover of Raman peak position shifts shown in **Figure 3d**. To further corroborate this conclusion, we compared the ratio of $Cu^O/Cu^I$ site occupancy derived from XRD and the peak intensity ratio of the two sites' contribution based on Raman spectrum. The two results are in good agreement with each other (**Figure 3e**).

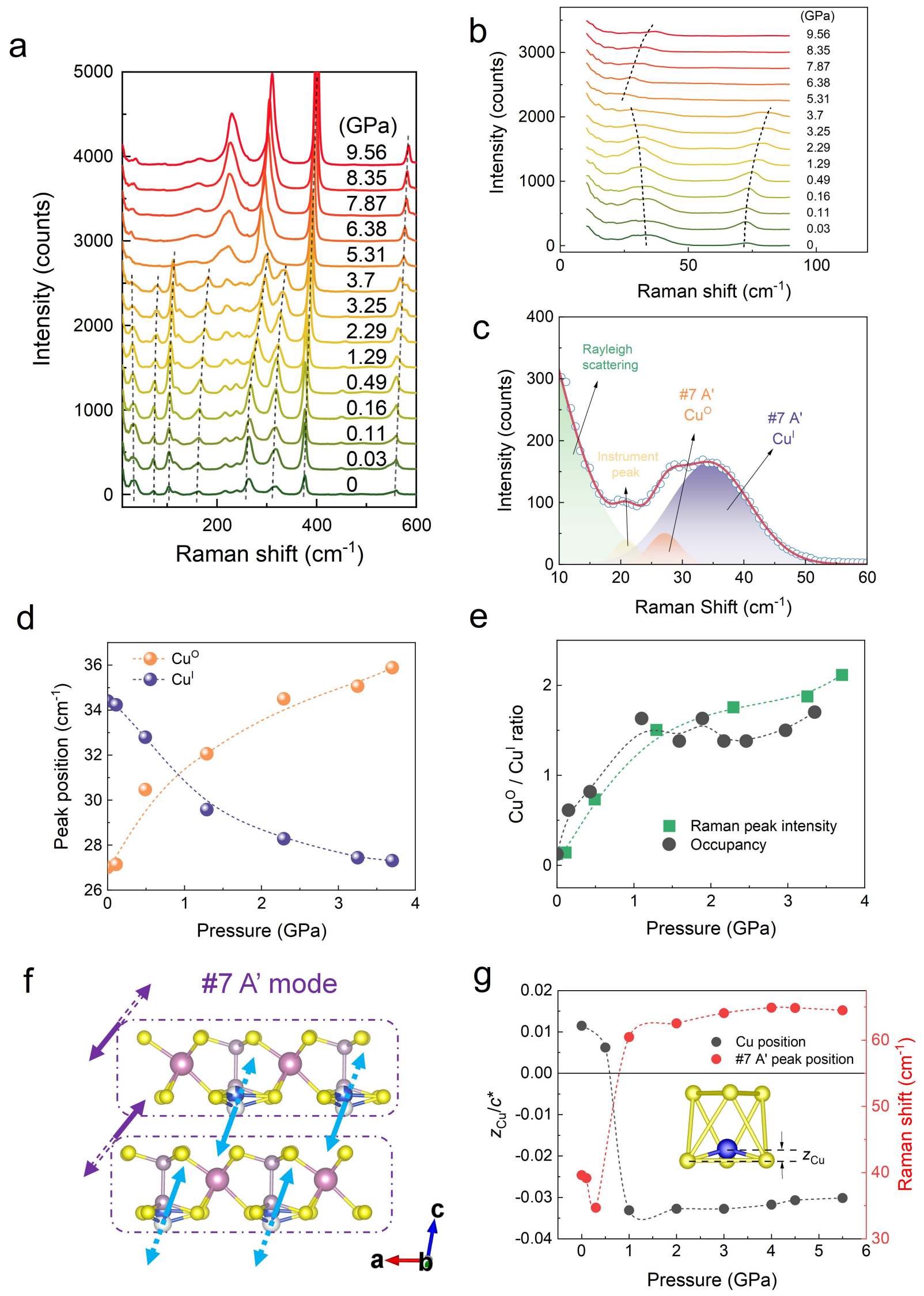

**Figure 3. Pressure-dependent vibrational spectra.** (a) Raman spectra of CIPS at different pressures. (b) Zoomed-in plot of the low-frequency part of the Raman spectra. (c) Detailed deconvolution of the low-frequency Raman peaks by Gaussian multi-peak fitting. The peak around 30 $cm^{-1}$ can be separated into two components due to the contribution from $Cu^O$ and $Cu^I$ sites. (d) The peak positions of the two sub-peaks as a function of the pressure. (e) A comparison between the $Cu^O/Cu^I$ occupancy ratio and the Raman peak intensity ratio as a function of the pressure. (f) A cartoon depicts the #7 A' mode. The blue arrows indicate the motions of Cu ions, while the purple ones denote the overall motions of each layer. (g) DFT calculated $z$ position of the Cu ion and corresponding peak position of the #7 mode as a function of the pressure.

## Discussion

We now reconcile the experimental (mostly at 300 K) and calculation results (at 0 K) by considering the temperature effect and discuss in detail the relationship between Cu displacement and interlayer sliding during the pressurization and phase transition.

At 0 K, Cu completely occupies the $Cu^{I}$ site at atmospheric pressure. Under compression, the vibrational mode of Cu in the z direction continues to soften, resulting in an isosymmetric phase transition of the material – Cu changes from occupying the $Cu^{I}$ site to completely occupying the $Cu^{O}$ site over 1.0 GPa. Because the $Cu^{O}$ has strong interaction with the adjacent sulphur S4, the z-direction vibration mode (#7 mode) of Cu, like other vibration modes, now continues to harden under subsequent compression. At about 5.8 GPa, a first-order phase transition driven by the interlayer sliding takes place, converting the crystal symmetry from monoclinic (*Cc*) to trigonal (*P*31*c*). Accompanying the phase transition is the movement of Cu back to the off-center position inside the sulphur octahedron. However, different from the $Cu^{I}$ site of the *Cc* phase, this off-center site is closer to the center of the octahedral cage. The crystal structure and Cu site of the high-pressure phase is analogous to $CuInP_2Se_6$, which somewhat hints the close structural connection between these homologous compounds[29].

At room temperature under atmospheric pressure, Cu ions show a broadening of its distribution towards the metastable $Cu^{O}$ site in the vdW gap due to thermally-activated hopping motion. Upon compression, the free energy of the $Cu^{O}$ site is reduced (**Figure 4b**), driving more Cu to displace from the intra-layer $Cu^{I}$ site towards the in-gap $Cu^{O}$ site. The occupancy of these two sites thus depends delicately on the balance between their relative energy landscape and the thermally-activated hopping effect. As a result, a complete occupancy of the $Cu^{O}$ site was not seen at room temperature up to the highest pressure before phase transition. The conclusion is also confirmed by ferroelectric polarization measurements, where the polarization is in good agreement with the value calculated from a mixed occupancy of $Cu^{I}$ (low polarization, ~3.3 $\mu C/cm^2$) and $Cu^{O}$ (high polarization, ~12.2 $\mu C/cm^2$) sites[12] (**Figure S5d**). In conjunction with the polarization increase is the drastic enhancement of $T_C$ under compression, consistent with the negative piezoelectricity of CIPS. For an order-disorder-type ferroelectric like CIPS, the $T_C$ increment can be interpreted as the increase of the potential barrier for thermal hopping of Cu ions between two centrosymmetric off-center sites. Such explanation is also supported by the rise of activation energy for ionic conduction under increased pressure. The weak softening of the #7 zone-center mode can also be attributed to the competition behavior of the $Cu^{O}$ and $Cu^{I}$ sites with regard to pressure. Interestingly, #7 mode involves the out-of-plane displacive motion of the Cu ions in association with the shear motion of the entire layer (**Supplementary gif images**), suggesting the interlock coupling between Cu displacement and interlayer sliding as evidenced by the structural refinement results. In fact, through a group-subgroup analysis, it can be found that the *P*31*c* phase is transformed into *Cc* phase by two irreducible representations, $\Gamma_1^+$ and $\Gamma_3^-$ (**SI, gif image**), which involve out-of-plane motion of Cu and interlayer shearing, respectively. At the phase transition point, relative interlayer sliding by -$1/3\vec{a}$ along the ***a***-axis happens. Such a structural transformation from monoclinic to trigonal phase seems to be a universal behavior of metal thiophosphates [30-34], in which high lattice symmetry and more compact atomic packing possibly reduce the total energy under compressive pressure.

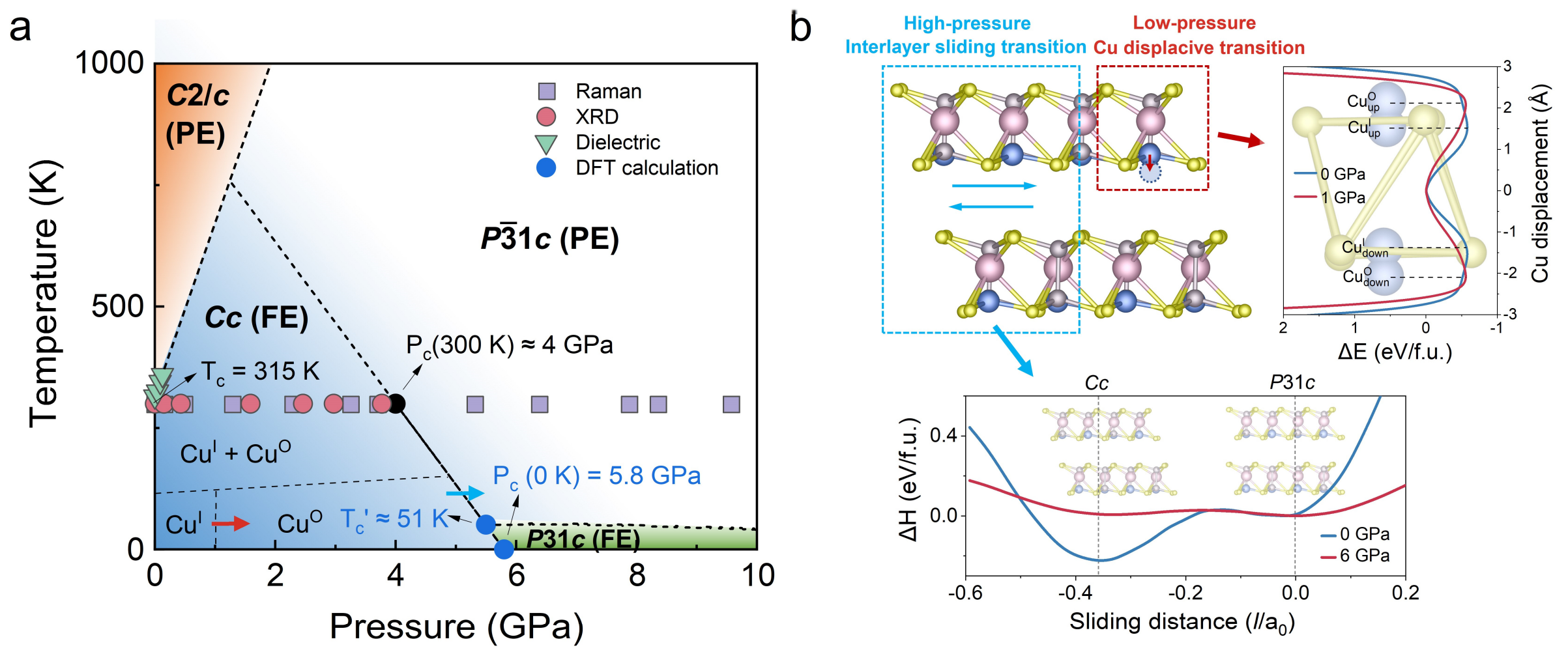

**Figure 4. Phase diagram and energy landscape in CIPS.** (a) Pressure-temperature (P - T) phase diagram of CIPS constructed jointly by experimental and theoretical results. The dash-line phase boundaries are qualitatively determined by connection and extrapolation among the data points. The red and blue arrows denote two different structural transition as detailed in (b). PE: paraelectric, FE: ferroelectric, $T_C$: Curie temperature, $P_C$: transition pressure. (b) Schematic energy landscapes of Cu displacive transition and interlayer sliding transition tuned by hydrostatic pressure. The sliding distance *l* is defined as the in-plane interlayer displacement between two In ions from nearest layers. Here, *l* is normalized to the in-plane lattice constant $a_0$ of $CuInP_2S_6$, such that $l/a_0 = 0$ ($l/a_0 = 1/3$) explicitly corresponds to the *P*31*c* (*Cc*) phase.

In the case of CIPS, the interlayer sliding causes the loss of bonding between Cu and the adjacent S4 atom. Consequently, the Cu ion at the $Cu^O$ site becomes unstable, driving it back into the sulphur octahedral cage. In contrast to the polar phase (*P*31*c*) predicted at 0 K, we observed a non-polar phase (*P*$\bar{3}$1*c*) at room temperature due to the disordered distribution of Cu among two off-center sites and the central site, which is, again, a reminiscence of the paraelectric phase of $CuInP_2Se_6$ [29]. This suggests there is a ferroelectric-paraelectric phase transition at finite temperature for the high-pressure $P31c$ phase of CIPS. The Curie temperature of $P31c$ phase was estimated through Landau φ4 potential, while Ginzburg term was not considered based on previous results on $Cc$ phase [35]. Through analyzing its Boltzmann factor ($A_0e^{-Ep/kT}$), we determine that the $T_C$ of the $P31c$ phase is about 51 K. Here, the calculated energy barrier $E_p$ between the ferroelectric $P31c$ and non-polar $P\bar{3}1c$ phase is ~ 28.7 meV/f.u. To simplify, the constant $A_0$ is estimated by fitting bulk $Cc$ phase ($T_C$ ~ 225 K, $E_p$ ~ 120 meV/f.u.). By summarizing the experimental and theoretical findings, a pressure-temperature phase diagram of CIPS can be constructed as shown in **Figure 4a**. The phase diagram highlights the effect of pressure on the energy landscapes of Cu displacive transition in low-pressure region and interlayer sliding transition in high-pressure region. As shown in **Figure 4b**, the red box represents the displacive transition of the Cu ions in low-pressure region, and its corresponding energy profile change is shown in the right panel of the figure. Our NEB calculations reveal that under ambient pressure (0 GPa), Cu ions prefer to stay inside the sulfur layer. However, under small pressure conditions (e.g., 1 GPa), Cu ions tend to move towards the vdW gap, as schematically shown in red box of **Figure 4b**. On the other hand, the blue box represents the interlayer sliding transition of the whole layer, revealing an asymmetrical double-well potential in the sliding potential curve (the bottom panel of **Figure 4b**). Specifically, the enthalpy of the $Cc$ phase is lower than that of the $P31c$ phase ($H_{Cc}$- $H_{P31c}$ = -0.21 eV/f.u.) under ambient pressure (0 GPa). Remarkably, as pressure increases beyond the transition point, for instance, enthalpy difference $H_{Cc}$- $H_{P31c}$ = 0.05 eV/f.u. at 6 GPa, the $P31c$ phase becomes more stable than the $Cc$ phase. Furthermore, pressure has the effect of reducing the sliding barrier between the $P31c$ and $Cc$ phases, promoting the occurrence of the sliding transition. Importantly, all theoretical results are in excellent agreement with experimental findings. Additionally, our DFT calculation results indicate that the sign of the longitudinal piezoelectric coefficient ($d_{33}$) is directly associated with the Cu position in the lattice. The calculated $d_{33}$ for the $Cc$ phase with total $Cu^I$ or $Cu^O$ occupancy is -7.01 pC/N and 1.6 pC/N, respectively, consistent with previous report [15]. The value changes to 7.08 pC/N for the $P31c$ phase. This finding signifies the importance of stacking order on the electromechanical behavior of the crystal lattice.

We note that a related report[36] was recently published during the preparation of our manuscript. However, the report by Yao et al. focused mainly on the polarization enhancement in the early stage of the pressurization, which is part of our results reported herein. Moreover, different from their report, we observed a continuous increase of polarization after 1.5 GPa, which is consistent with our structural results and previous study by second harmonic generation[24]. This results in a totally different interpretation of the Cu displacement behavior after 1.5 GPa.

## Conclusion

Through a comprehensive analysis of combined experimental and theoretical results, this work unravels the complicated yet intriguing structural transformation of vdW layered ferroelectric $CuInP_2S_6$ under hydrostatic pressure. In low-pressure region (< 0.5 GPa), the compression renders the $Cu^O$ site energetically more favorable, thus resulting in an abrupt increase of the $Cu^O$ site occupancy and consequently the polarization enhancement. However, due to the small energy difference between these two sites, a full occupancy of $Cu^O$ at room temperature is not realized by further pressurization due to thermally-assisted hopping, prior to the structural phase transition driven by interlayer sliding. A soft mode with combined Cu translation and interlayer shear motions is identified to closely link with this sliding-mediated phase transition. The pressure-induced structural evolution of CIPS pivots on the displacive instability of Cu ion, which serves as a bridging switch to modulate the interlayer interaction in response to external pressure. The interlayer sliding induced changes of crystal symmetry in conjunction with the ferroelectric polarization highlights again the critical role of interlayer stacking pattern in tailoring the long-range order parameters in vdW layer crystals.

# Methods

## Materials synthesis

$CuInP_2S_6$ single crystals were synthesized by chemical vapor transport (CVT) method as reported previously [13,37], with iodine as the transport agent. The composition stoichiometry and phase purity were checked by energy dispersive X-ray spectroscopy (EDS) and Raman spectroscopy. High purity and regular shape crystals are selected for subsequent measurements.

## Single-crystal X-ray Diffraction under high pressure

A pale yellow, platelet single crystal of $CuInP_2S_6$ characterized under ambient conditions was loaded into a Merrill-Bassett diamond anvil cell (DAC) [38], equipped with Boehlar-Almax diamond anvils, tungsten carbide backing seats [39] (opening angle = 40°) and a tungsten gasket. The sample chamber was filled with a 4:1 volumetric mixture of methanol:ethanol to serve as a pressure-transmitting medium, and the pressure within the sample chamber was measured using the ruby fluorescence method [40].

The crystal structure of $CuInP_2S_6$ was characterized under ambient conditions of temperature and pressure using a Rigaku Synergy-S diffractometer with mirror-chromated Mo Kα radiation ($\lambda$ = 0.7107 Å). Diffraction data were integrated and reduced in CrysAlis Pro[41] and an absorption correction was applied using SADABS [42]. The crystal structure was solved using ShelXT [43] and refined against $|F^2|$ using ShelXL [44] in Olex2 [45] as a two-component inversion twin. All geometric and anisotropic displacement parameters were refined freely. The detailed procedures of data collection and structural refinement are described in **Supplementary Text 2**. The refined crystallographic data are included in **Table S2** and **S3**.

## Electrical measurements under high pressure

$CuInP_2S_6$ crystals with thicknesses of 10~20 μm were selected to fabricate the parallel-plate ferroelectric capacitors with Au electrodes. The sample was then loaded into a hydrostatic pressure cell (HPC-33, Quantum Design) for high-pressure electrical tests. The sample was connected in series to the external circuit through the sample board and sample feedthrough of the instrument. A hydraulic press set (Model FHP-5P-37S/40S) was used to apply pressure, and the pressurizing medium is daphne 7373 oil. The upper pressure and temperature limit of the instrument is 3.0 GPa and 400 K, respectively.

A commercial ferroelectric tester (Radiant Technologies, Multiferroic II) was used to measure the standard polarization hysteresis loop and remanent hysteresis loop. Dielectric data upon different temperature under each pressure were collected by a commercial LCR meter (Agilent, 4284A) in conjunction with a temperature chamber (Model 9023, Delta Design).

**Raman spectroscopy under high pressure**

Raman spectroscopies were conducted on a confocal Raman microscope (WITec Alpha 300RAS) in backscattering geometry with an excitation laser wavelength of 633 nm (He−Ne laser) and an objective lens (×50) with a long working distance and two notch filters. Laser power was kept as high as 8 mW with the laser spot size of 1 μm to obtain sufficient Raman signal. The high-pressure cell used in this experiment was a symmetric Mao-type DAC with two diamonds with 400 μm culet size. The sample and pressure-transmitting medium (silicone oil) with a small ruby piece (~10 μm, for pressure calibration) were loaded into a hole of 150 μm diameter drilled in a gasket (T301, 250 μm thick) that had been pre-indented.

**Computational method**

The density-functional theory (DFT) calculations were performed using the plane wave basis as implemented in Vienna *ab initio* Simulation Package (VASP) [46]. The generalized gradient approximation in the Perdew-Burke-Ernzerhof (GGA-PBE) formulation is adopted with a cutoff energy of 500 eV [47]. Brillouin zones are sampled using 6×6×3 grid *k*-points in the Monkhorst-Pack scheme. The phonon frequencies are determined using density functional perturbation theory (DFPT), which is implemented in the VASP code. The Raman tensor is calculated through macroscopic dielectric tensor [48]. The evaluation of polarization is conducted using the Berry-phase method [49], wherein both the ionic and electronic contributions are taken into account. The nudged elastic band (NEB) method is employed to generate and calculate a series of transition states between the ferroelectric and paraelectric states [50].

**Supplementary Material**

See the supplementary material for more information on the analyses of single-crystal XRD, dielectric and polarization measurements, Raman spectroscopy results and DFT calculations.

**Acknowledgements**

L.Y. and L.F. acknowledge the support by National Natural Science Foundation of China (No. 12074278), the Natural Science Foundation of the Jiangsu Higher Education Institution of China (20KJA140001), and the Priority Academic Program Development (PAPD) of Jiangsu Higher Education Institutions. L.Y. also acknowledges the support from Suzhou Science and Technology Bureau (ZXL2022514) and Jiangsu Specially-Appointed Professors Program. S.D. acknowledge the support by National Natural Science Foundation of China (No. 12274069). J.J.Z. acknowledges financial support by the Natural Science Foundation of the Jiangsu Province Grant No. BK20230806. J.J.Z. and S.D. also thank the Big Data Center of Southeast University for providing the computational resource. S.A.M. acknowledges the support of the Australian Research Council (ARC) from a Future Fellowship (FT200100243) and Discovery Project (DP220103690).

**AUTHOR DECLARATIONS**

**Conflict of Interest**

The authors have no conflicts to disclose.

**DATA AVAILABILITY**

The data that support the findings of this study are available from the corresponding authors upon reasonable request.

## References

1 C. Wang, L. You, D. Cobden, and J. Wang, Towards two-dimensional van der Waals ferroelectrics, Nat. Mater. **22**, 542 (2023).

2 D. Zhang, P. Schoenherr, P. Sharma, and J. Seidel, Ferroelectric order in van der Waals layered materials, Nat. Rev. Mater. **8**, 25 (2023).

3 M. Wu and J. Li, Sliding ferroelectricity in 2D van der Waals materials: Related physics and future opportunities, Proceedings of the National Academy of Sciences **118**, e2115703118 (2021).

4 Z. Zheng, Q. Ma, Z. Bi, S. de la Barrera, M.-H. Liu, N. Mao, Y. Zhang, N. Kiper, K. Watanabe, T. Taniguchi, J. Kong, W. A. Tisdale, R. Ashoori, N. Gedik, L. Fu, S.-Y. Xu, and P. Jarillo-Herrero, Unconventional ferroelectricity in moiré heterostructures, Nature **588**, 71 (2020).

5 M. V. Stern, Y. Waschitz, W. Cao, I. Nevo, K. Watanabe, T. Taniguchi, E. Sela, M. Urbakh, O. Hod, and M. B. Shalom, Interfacial ferroelectricity by van der Waals sliding, Science **372**, 1462 (2021).

6 K. Yasuda, X. Wang, K. Watanabe, T. Taniguchi, and P. Jarillo-Herrero, Stacking-engineered ferroelectricity in bilayer boron nitride, Science **372**, 1458 (2021).

7 S. Deb, W. Cao, N. Raab, K. Watanabe, T. Taniguchi, M. Goldstein, L. Kronik, M. Urbakh, O. Hod, and M. Ben Shalom, Cumulative polarization in conductive interfacial ferroelectrics, Nature **612**, 465 (2022).

8 P. Meng, Y. Wu, R. Bian, E. Pan, B. Dong, X. Zhao, J. Chen, L. Wu, Y. Sun, Q. Fu, Q. Liu, D. Shi, Q. Zhang, Y.-W. Zhang, Z. Liu, and F. Liu, Sliding induced multiple polarization states in two-dimensional ferroelectrics, Nat. Commun. **13**, 7696 (2022).

9 A. Belianinov, Q. He, A. Dziaugys, P. Maksymovych, E. Eliseev, A. Borisevich, A. Morozovska, J. Banys, Y. Vysochanskii, and S. V. Kalinin, CuInP2S6 Room Temperature Layered Ferroelectric, Nano Lett. **15**, 3808 (2015).

10 F. Liu, L. You, K. L. Seyler, X. Li, P. Yu, J. Lin, X. Wang, J. Zhou, H. Wang, H. He, S. T. Pantelides, W. Zhou, P. Sharma, X. Xu, P. M. Ajayan, J. Wang, and Z. Liu, Room-temperature ferroelectricity in CuInP2S6 ultrathin flakes, Nat. Commun. **7**, 12357 (2016).

11 S. M. Neumayer, E. A. Eliseev, M. A. Susner, A. Tselev, B. J. Rodriguez, J. A. Brehm, S. T. Pantelides, G. Panchapakesan, S. Jesse, S. V. Kalinin, M. A. McGuire, A. N. Morozovska, P. Maksymovych, and N. Balke, Giant negative electrostriction and dielectric tunability in a van der Waals layered ferroelectric, Phys. Rev. Mater. **3**, 024401 (2019).

12 L. You, Y. Zhang, S. Zhou, A. Chaturvedi, S. A. Morris, F. Liu, L. Chang, D. Ichinose, H. Funakubo, W. Hu, T. Wu, Z. Liu, S. Dong, and J. Wang, Origin of giant negative piezoelectricity in a layered van der Waals ferroelectric, Sci. Adv. **5**, eaav3780 (2019).

13 S. Zhou, L. You, A. Chaturvedi, S. A. Morris, J. S. Herrin, N. Zhang, A. Abdelsamie, Y. Hu, J. Chen, Y. Zhou, S. Dong, and J. Wang, Anomalous polarization switching and permanent retention in a ferroelectric ionic conductor, Mater. Horiz. **7**, 263 (2020).

14 S. M. Neumayer, M. Si, J. Li, P.-Y. Liao, L. Tao, A. O'Hara, S. T. Pantelides, P. D. Ye, P. Maksymovych, and N. Balke, Ionic Control over Ferroelectricity in 2D Layered van der Waals Capacitors, ACS Appl. Mat. Interfaces **14**, 3018 (2022).

15 J. A. Brehm, S. M. Neumayer, L. Tao, A. O'Hara, M. Chyasnavichus, M. A. Susner, M. A. McGuire, S. V. Kalinin, S. Jesse, P. Ganesh, S. T. Pantelides, P. Maksymovych, and N. Balke,

Tunable quadruple-well ferroelectric van der Waals crystals, Nat. Mater. **19**, 43 (2020).
16 G. A. Samara, T. Sakudo, and K. Yoshimitsu, Important Generalization Concerning the Role of Competing Forces in Displacive Phase Transitions, Phys. Rev. Lett. **35**, 1767 (1975).
17 G. A. Samara, in *Solid State Physics*, edited by Henry Ehrenreich and Frans Spaepen (Academic Press, 2001), Vol. 56, pp. 239.
18 A. Grzechnik, V. B. Cajipe, C. Payen, and P. F. McMillan, Pressure-induced phase transition in ferrielectric CuInP2S6, Solid State Commun. **108**, 43 (1998).
19 P. Guranich, V. Shusta, E. Gerzanich, A. Slivka, I. Kuritsa, and O. Gomonnai, Influence of hydrostatic pressure on the dielectric properties of CuInP2S6 and CuInP2Se6 layered crystals, J. Phys: Conf. Ser. **79**, (2007).
20 P. P. Guranich, A. G. Slivka, V. S. Shusta, O. O. Gomonnai, and I. P. Prits, Optical and dielectric properties of CuInP2S6layered crystals at high hydrostatic pressure, J. Phys: Conf. Ser. **121**, (2008).
21 Q. He, A. Belianinov, A. Dziaugys, P. Maksymovych, Y. Vysochanskii, S. V. Kalinin, and A. Y. Borisevich, Antisite defects in layered multiferroic CuCr0.9In0.1P2S6, Nanoscale **7**, 18579 (2015).
22 V. Maisonneuve, V. B. Cajipe, A. Simon, R. Von Der Muhll, and J. Ravez, Ferrielectric ordering in lamellar $CuInP_2S_6$, Phys. Rev. B **56**, 10860 (1997).
23 S.-H. Wei, S. B. Zhang, and A. Zunger, Off-center atomic displacements in zinc-blende semiconductor, Phys. Rev. Lett. **70**, 1639 (1993).
24 K. Bu, T. Fu, Z. Du, X. Feng, D. Wang, Z. Li, S. Guo, Z. Sun, H. Luo, G. Liu, Y. Ding, T. Zhai, Q. Li, and X. Lü, Enhanced Second-Harmonic Generation of van der Waals CuInP2S6 via Pressure-Regulated Cationic Displacement, Chem. Mater. **35**, 242 (2023).
25 S. N. Neal, S. Singh, X. Fang, C. Won, F.-t. Huang, S.-W. Cheong, K. M. Rabe, D. Vanderbilt, and J. L. Musfeldt, Vibrational properties of $CuInP_2S_6$ across the ferroelectric transition, Phys. Rev. B **105**, 075151 (2022).
26 A. K. Jonscher, The 'universal' dielectric response, Nature **267**, 673 (1977).
27 K. J. Laidler, The development of the Arrhenius equation, J. Chem. Educ. **61**, 494 (1984).
28 Y. M. Vysochanskii, V. A. Stephanovich, A. A. Molnar, V. B. Cajipe, and X. Bourdon, Raman spectroscopy study of the ferrielectric-paraelectric transition in layeredCuInP2S6, Phys. Rev. B **58**, 9119 (1998).
29 A. Dziaugys, K. Kelley, J. A. Brehm, L. Tao, A. Puretzky, T. Feng, A. O'Hara, S. Neumayer, M. Chyasnavichyus, E. A. Eliseev, J. Banys, Y. Vysochanskii, F. Ye, B. C. Chakoumakos, M. A. Susner, M. A. McGuire, S. V. Kalinin, P. Ganesh, N. Balke, S. T. Pantelides, A. N. Morozovska, and P. Maksymovych, Piezoelectric domain walls in van der Waals antiferroelectric CuInP2Se6, Nat. Commun. **11**, 3623 (2020).
30 C. R. S. Haines, M. J. Coak, A. R. Wildes, G. I. Lampronti, C. Liu, P. Nahai-Williamson, H. Hamidov, D. Daisenberger, and S. S. Saxena, Pressure-Induced Electronic and Structural Phase Evolution in the van der Waals Compound $FePS_3$, Phys. Rev. Lett. **121**, 266801 (2018).
31 N. C. Harms, H.-S. Kim, A. J. Clune, K. A. Smith, K. R. O'Neal, A. V. Haglund, D. G. Mandrus, Z. Liu, K. Haule, D. Vanderbilt, and J. L. Musfeldt, Piezochromism in the magnetic chalcogenide MnPS3, npj Quantum Materials **5**, 56 (2020).
32 M. J. Coak, D. M. Jarvis, H. Hamidov, A. R. Wildes, J. A. M. Paddison, C. Liu, C. R. S. Haines, N. T. Dang, S. E. Kichanov, B. N. Savenko, S. Lee, M. Kratochvílová, S. Klotz, T. C. Hansen,

D. P. Kozlenko, J.-G. Park, and S. S. Saxena, Emergent Magnetic Phases in Pressure-Tuned van der Waals Antiferromagnet $FePS_3$, Physical Review X **11**, 011024 (2021).
33 N. C. Harms, T. Matsuoka, S. Samanta, A. J. Clune, K. A. Smith, A. V. Haglund, E. Feng, H. Cao, J. S. Smith, D. G. Mandrus, H.-S. Kim, Z. Liu, and J. L. Musfeldt, Symmetry progression and possible polar metallicity in NiPS3 under pressure, npj 2D Materials and Applications **6**, 40 (2022).
34 T. Matsuoka, R. Rao, M. A. Susner, B. S. Conner, D. Zhang, and D. Mandrus, Pressure-induced insulator-to-metal transition in the van der Waals compound $CoPS_3$, Phys. Rev. B **107**, 165125 (2023).
35 W. Song, R. Fei, and L. Yang, Off-plane polarization ordering in metal chalcogen diphosphates from bulk to monolayer, Phys. Rev. B **96**, 235420 (2017).
36 X. Yao, Y. Bai, C. Jin, X. Zhang, Q. Zheng, Z. Xu, L. Chen, S. Wang, Y. Liu, J. Wang, and J. Zhu, Anomalous polarization enhancement in a van der Waals ferroelectric material under pressure, Nat. Commun. **14**, 4301 (2023).
37 F. Liu, L. You, K. L. Seyler, X. Li, P. Yu, J. Lin, X. Wang, J. Zhou, H. Wang, H. He, S. T. Pantelides, W. Zhou, P. Sharma, X. Xu, P. M. Ajayan, J. Wang, and Z. Liu, Room-temperature ferroelectricity in CuInP2S6 ultrathin flakes, Nat Commun **7**, 12357 (2016).
38 L. Merrill and W. A. Bassett, Miniature diamond anvil pressure cell for single crystal x-ray diffraction studies, Rev. Sci. Instrum. **45**, 290 (1974).
39 S. A. Moggach, D. R. Allan, S. Parsons, and J. E. Warren, Incorporation of a new design of backing seat and anvil in a Merrill–Bassett diamond anvil cell, J. Appl. Cryst. **41**, 249 (2008).
40 G. J. Piermarini, S. Block, J. Barnett, and R. Forman, Calibration of the pressure dependence of the R 1 ruby fluorescence line to 195 kbar, J. Appl. Phys. **46**, 2774 (1975).
41 R. O. Diffraction, CrysAlisPro Software system, version 1.171. 40.67 a (Rigaku Corporation Wroclaw, Poland, 2019).
42 G. Sheldrick, SADABS Bruker Axs Inc, Madison, Wisconsin, USA (2007).
43 G. M. Sheldrick, SHELXT–Integrated space-group and crystal-structure determination, Acta Cryst A **71**, 3 (2015).
44 G. M. Sheldrick, Crystal structure refinement with SHELXL, Acta Cryst C **71**, 3 (2015).
45 O. V. Dolomanov, L. J. Bourhis, R. J. Gildea, J. A. K. Howard, and H. Puschmann, OLEX2: a complete structure solution, refinement and analysis program, J Appl Cryst **42**, 339 (2009).
46 G. Kresse and J. Furthmüller, Efficient iterative schemes for ab initio total-energy calculations using a plane-wave basis set, Phys. Rev. B **54**, 11169 (1996).
47 J. P. Perdew, K. Burke, and M. Ernzerhof, Generalized Gradient Approximation Made Simple, Phys. Rev. Lett. **77**, 3865 (1996).
48 D. Porezag and M. R. Pederson, Infrared intensities and Raman-scattering activities within density-functional theory, Phys. Rev. B **54**, 7830 (1996).
49 R. Resta, M. Posternak, and A. Baldereschi, Towards a quantum theory of polarization in ferroelectrics: The case of $KNbO_3$, Phys. Rev. Lett. **70**, 1010 (1993).
50 G. Henkelman, B. P. Uberuaga, and H. Jónsson, A climbing image nudged elastic band method for finding saddle points and minimum energy paths, The Journal of Chemical Physics **113**, 9901 (2000).

**Supplementary Online Material**

**Sliding-mediated ferroelectric phase transition in $CuInP_2S_6$ under pressure**

Zhou Zhou[1,#], Jun-Jie Zhang[2,#], Gemma F. Turner[3,#], Stephen A. Moggach[3,*], Yulia Lekina[4], Samuel Morris[5], Shun Wang[1], Yiqi Hu[1], Qiankun Li[1], Jinshuo Xue[1], Zhijian Feng[1], Qingyu Yan[1], Yuyan Weng[1], Bin Xu[1], Yong Fang[6], Ze Xiang Shen[4], Liang Fang[1,*], Shuai Dong[2,*], Lu You[1,*]

[1]School of Physical Science and Technology, Jiangsu Key Laboratory of Thin Films, Soochow University, Suzhou, 215006, China

[2]Key Laboratory of Quantum Materials and Devices of Ministry of Education, School of Physics, Southeast University, Nanjing 211189, China

[3]School of Molecular Sciences, The University of Western Australia, 35 Stirling Highway, Crawley, Perth, Western Australia 6009, Australia.

[4]Centre for Disruptive Photonic Technologies, Nanyang Technological University, 637371, Singapore

[5]Facility for Analysis, Characterisation, Testing and Simulation (FACTS), Nanyang Technological University, 639798, Singapore

[6]Jiangsu Laboratory of Advanced Functional Materials, School of Electronic and Information Engineering, Changshu Institute of Technology, Changshu, 215500, China

#These authors contributed equally.

*Corresponding authors:

stephen.moggach@uwa.edu.au

lfang@suda.edu.cn

sdong@seu.edu.cn

lyou@suda.edu.cn

**Supplementary Figure S1**

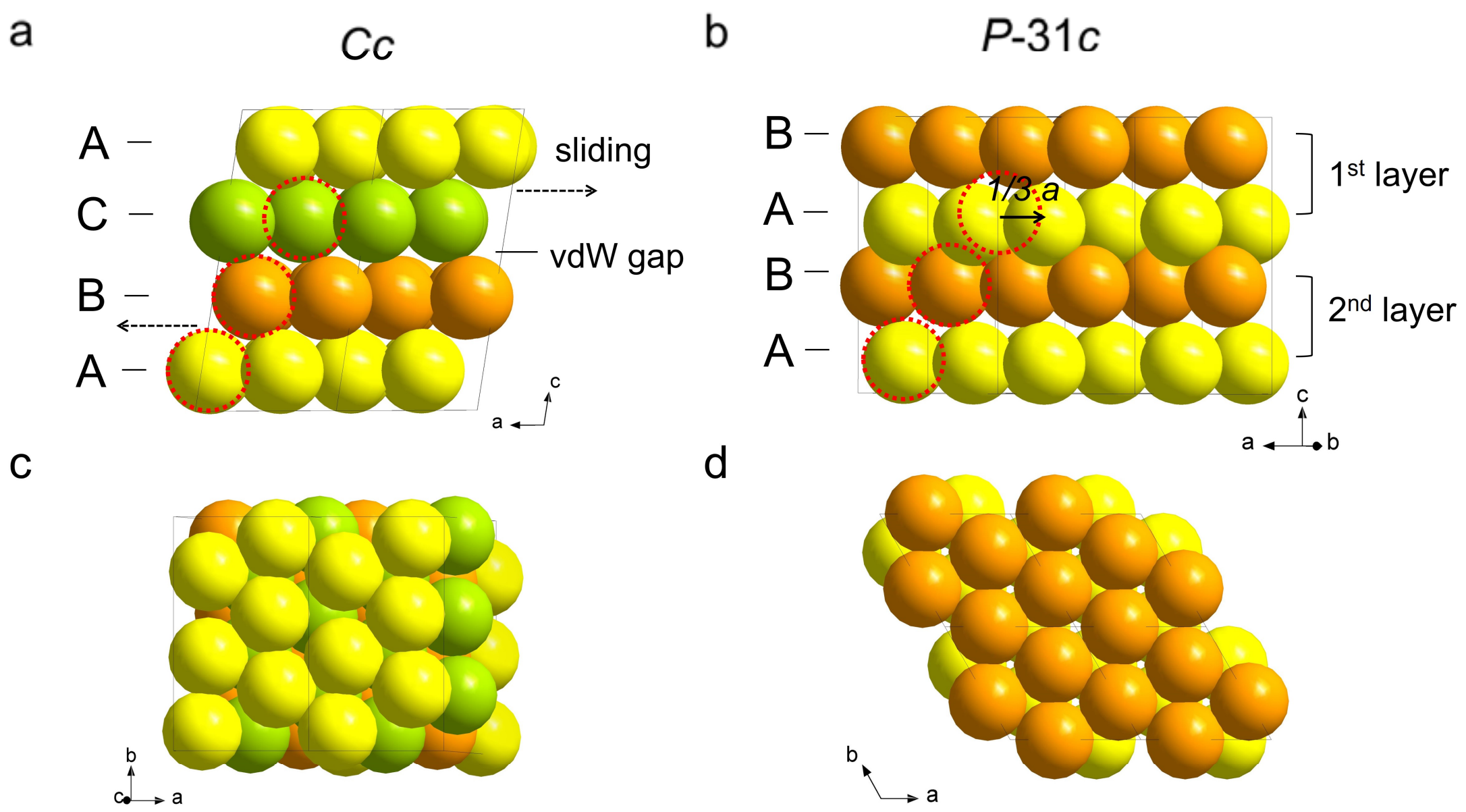

**Figure S1. Stacking patterns of sulphur atoms for different CIPS phases.** (a) Monoclinic *Cc* phase viewed along the *b*-axis and (c) perpendicular to *ab* plane. (b) Trigonal *P*-31*c* phase viewed perpendicular to *ac* plane and (d) along the *c*-axis. The viewing directions are intentionally selected for direct comparison of the atomic stacking between the two phases. Dotted circles indicate the original positions of tri-layer sulphur atoms in the *Cc* phase. In the *P*-31*c* phase, the top sulphur atom shifts by $1/3\vec{a}$ along the *a*-axis of *Cc* phase, changing the stacking pattern from ABC- to AB-type.

**Supplementary Figure S2**

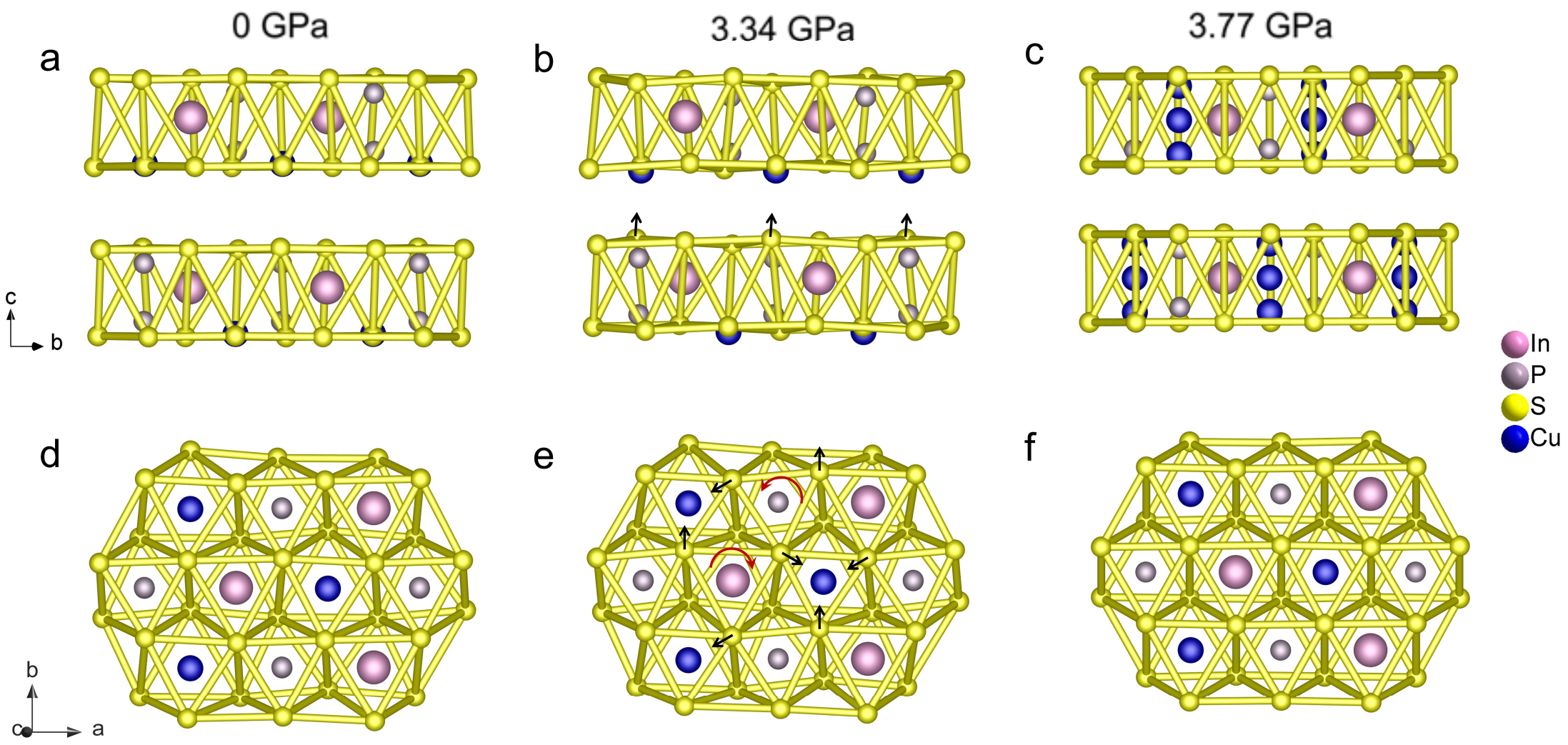

**Figure S2. Crystal structure of CIPS at different pressures.** (a, b, c) The atomic structure viewed along *a* axis of *Cc* phase at 0 (*Cc*), 3.34 (*Cc*), and 3.77 GPa (*P*-31*c*), respectively. (d, e, f) The atomic structure viewed perpendicular to a*b* plane at 0 (*Cc*), 3.34 (*Cc*), and 3.77 GPa (*P*-31*c*), respectively. The arrows in (b) denote the displacement direction of the S4 atoms towards the interlayer Cu atoms, which induces strong puckering of the sulphur layer and compression of the vdW gap. The black arrows in (e) indicate the in-plane displacements of the sulphur atoms due to the motion of Cu atoms from inside to outside of the S6 octahedra. This tendency causes the expansion and contraction of the two opposite S3 trigonal surfaces of the sulphur octahedra, as well as the clockwise and anti-clockwise rotations of the In and P-P sublattice octahedra, as indicated by the red arrows.

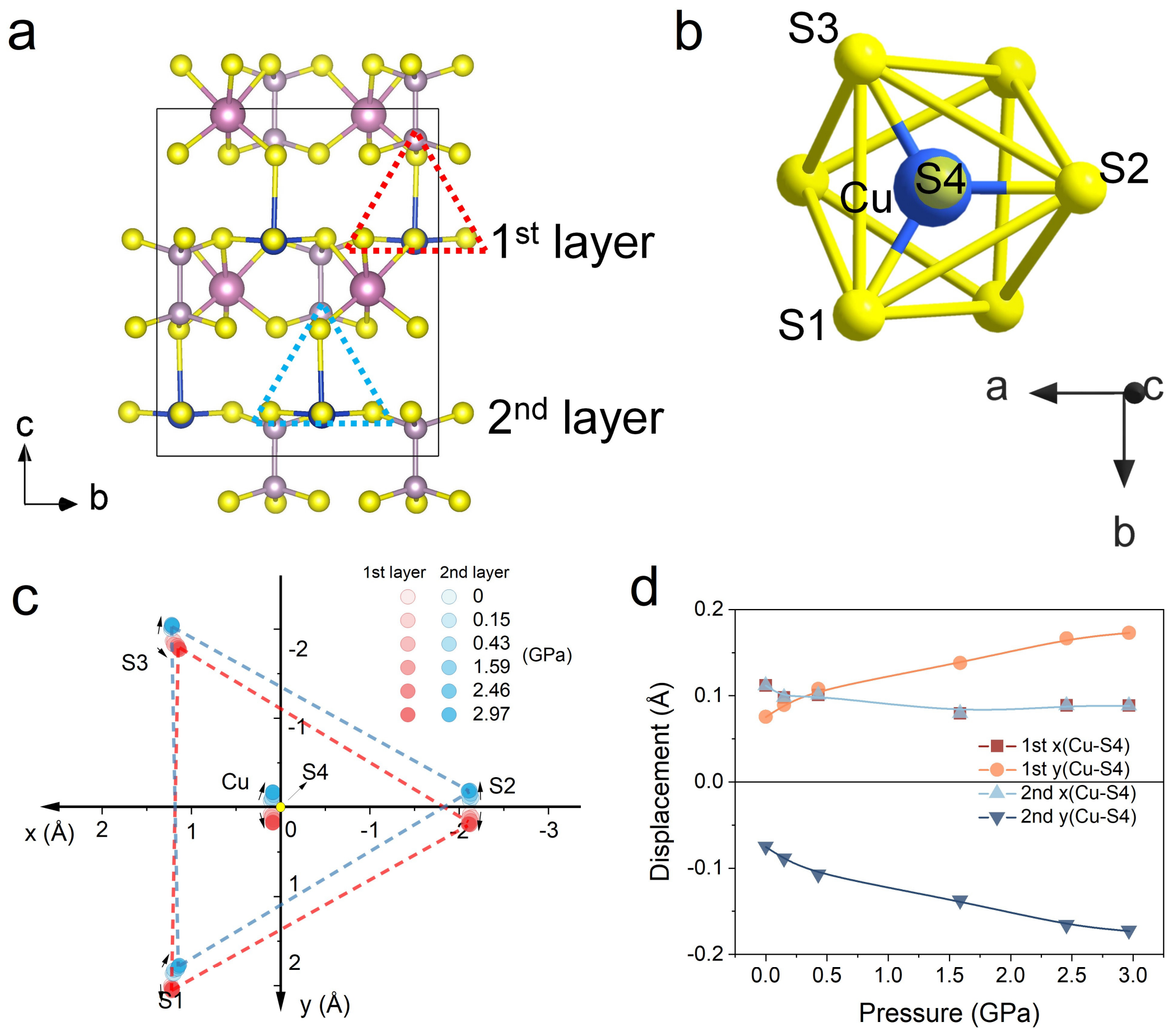

**Figure S3. In-plane displacements of Cu sublattice under pressure.** (a) Crystal structure of CIPS viewed along *a* axis. The color triangles denote the Cu-S tetrahedral sublattices of the bilayer unit cell. (b) Vertical view of the Cu-S sublattice. The Cu atom sits on top of the sulphur octahedron, and bonds with the topmost S4 atom across the vdW gap to form tetrahedral coordination. (c) Relative in-plane (xy/ab plane) displacements of the Cu and S atoms under different pressures with respect to S4 atom (origin). (d) Relative in-plane displacements of Cu with regard to S4 atom as a function of pressure.

## Supplementary Figure S4

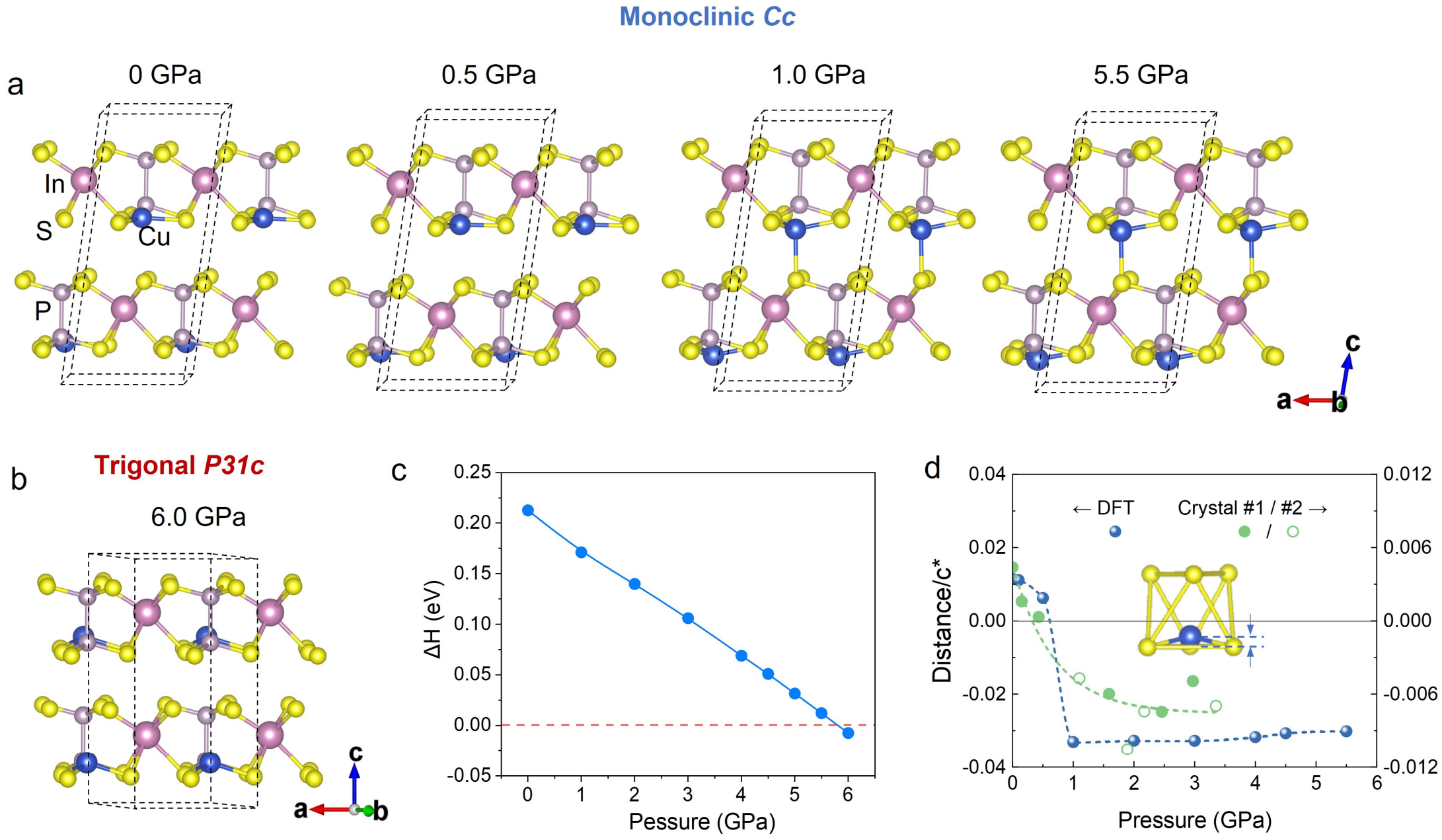

**Figure S4. Structural evolution by DFT calculations.** (a) Structures of CIPS *Cc* phase under different pressures from DFT calculations. (b) DFT calculated structure of CIPS $P31c$ phase. (c) Calculated enthalpy difference between *Cc* and $P\bar{3}1c$ phase as a function of pressure. (d) Comparison of Cu position variations in CIPS under different pressure from theoretical (blue) and experimental data (green).

## Supplementary Figure S5

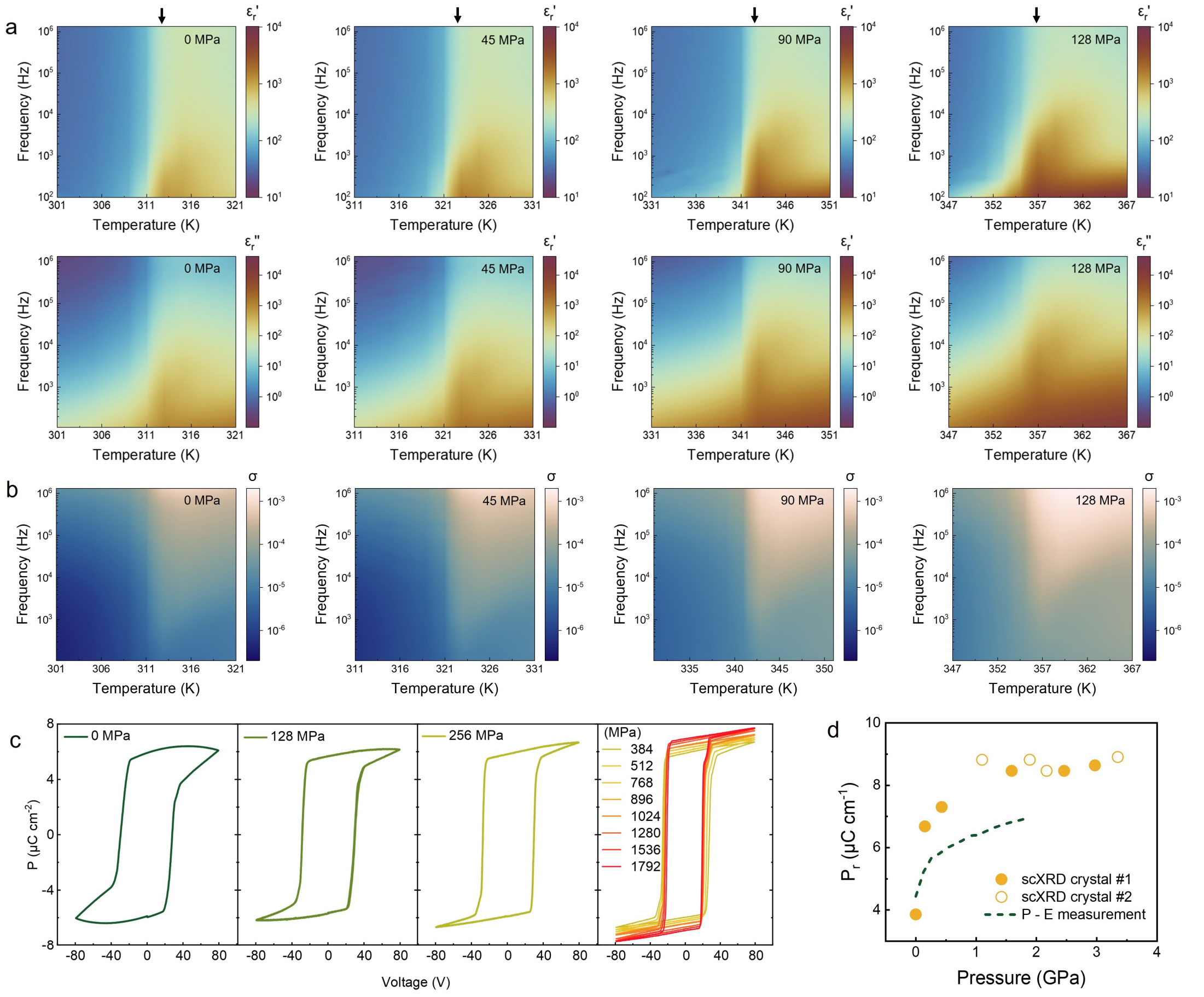

**Figure S5. Supplementary data series of dielectric and polarization measurements.** (a) Real (top row) and imaginary (bottom row) parts of relative dielectric permittivity and (b) AC conductivity as a function of frequency and temperature under 0, 45, 90, 128 MPa, respectively. The arrows indicate the ferroelectric-paraelectric transition temperature. (c) Standard polarization hysteresis loops of CIPS crystal measured at different pressures. The measurements were carried out at room temperature with a triangular voltage waveform of 12.5 Hz. (d) Comparison of ferroelectric polarization derived from direct P - E loop measurements (dash line) and those calculated based on mixed occupancies of $Cu^{I}$ (low polarization, ~3.3 µC/cm2) and $Cu^{O}$ (high polarization, ~12.2 µC/cm2) sites determined by scXRD refinements.

## Supplementary Figure S6

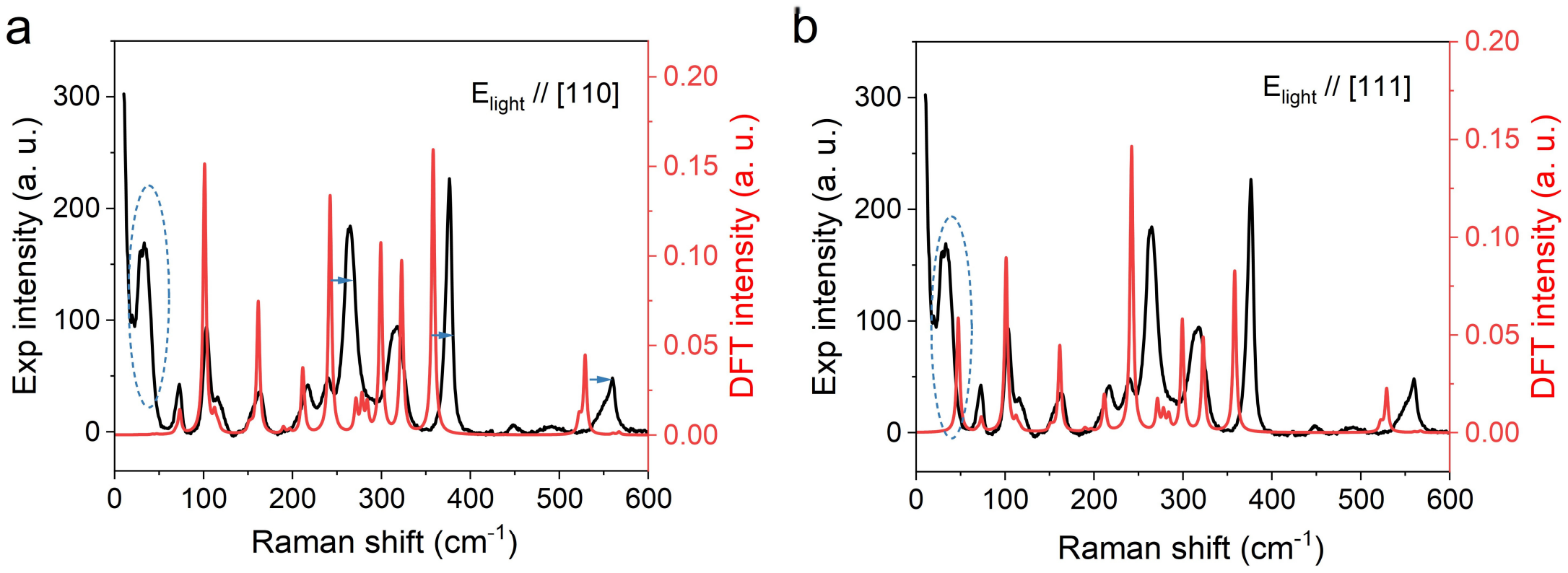

**Figure S6. Comparison of experimental and simulated Raman spectrum.** (a) Simulated Raman spectrum with the light polarization parallel to [110] direction. (b) Simulated Raman spectrum with the light polarization parallel to [111] direction. The calculation details of the Raman spectra from calculated Raman tensors (Table S1) can be found in Supplementary text 1.

## Supplementary Figure S7

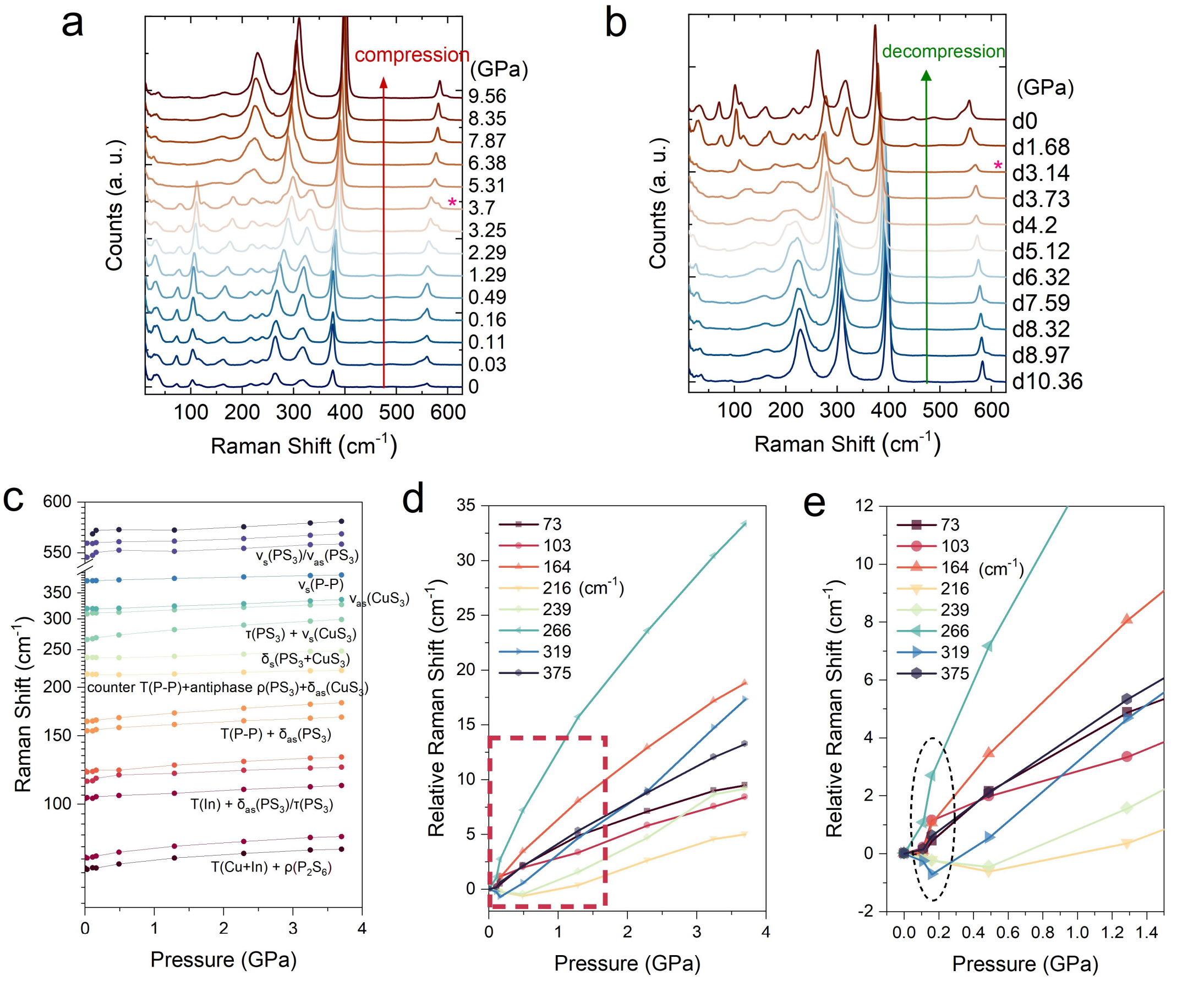

**Figure S7. Supplementary Raman spectra data.** (a) Compression and (b) decompression data series of the Raman spectra for CIPS single crystal. (c) Raman peak positions of different phonon modes as a function of pressure for *Cc* phase. T: translation, ρ: rocking, δ: bending, τ: twisting, ν: stretching, subscript s or as: symmetric or asymmetric. The readers are referred to the supplementary gif images of selected vibrational phonons for details. (d) Relative peak shifts of selected phonon modes as a function of pressure. (e) Zoomed-in image of the low-pressure part denoted by the dash box in (d).

## Supplementary Figure S8

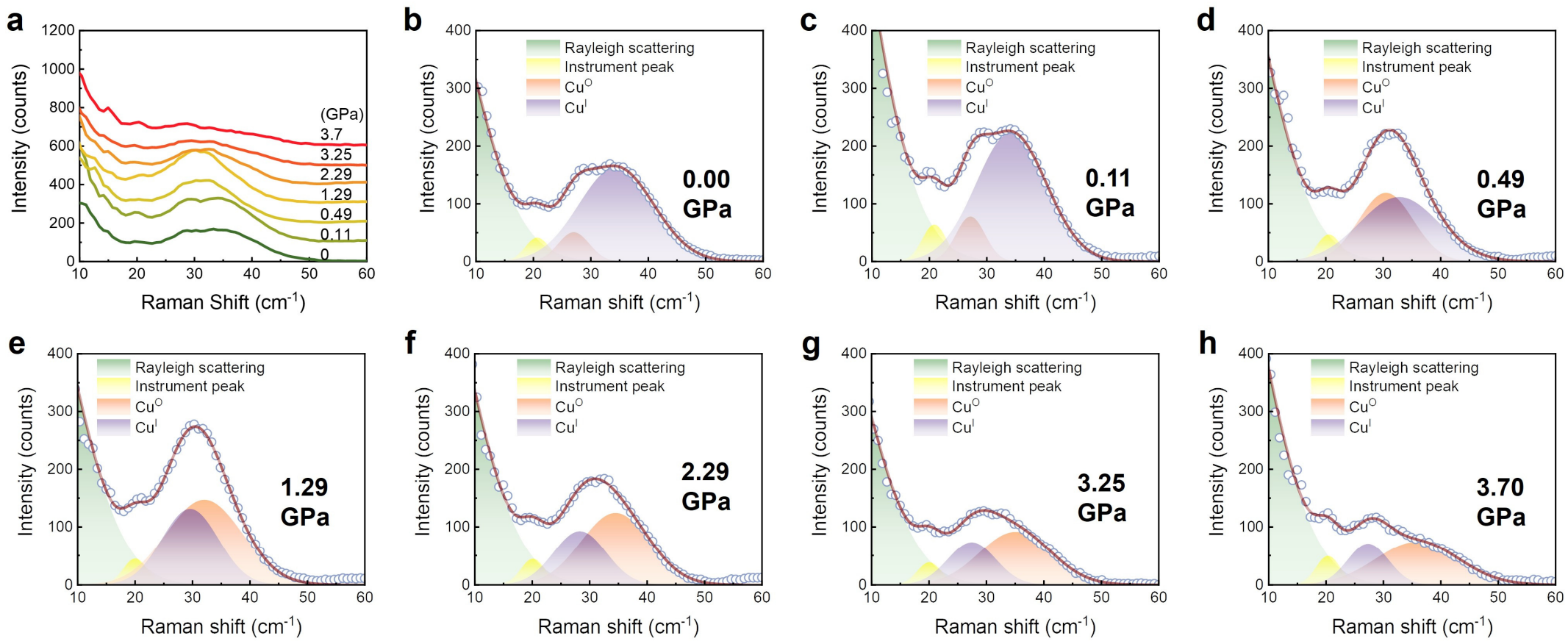

**Figure S8. Peak deconvolution of the low-frequency vibrational mode under different pressures.**

**Supplementary Text 1**

The intensity of Raman modes can be described by the equation(1) $I \propto \left|\overrightarrow{e_i}\boldsymbol{R}\overrightarrow{e_s}^T\right|^2$, where $\boldsymbol{R}$ is the Raman tensor, $\overrightarrow{e_i}$ and $\overrightarrow{e_s}$ represent the polarization vectors of incident and scattered light, respectively. Based on the point group (Class *m*) of *Cc* symmetry, Raman tensor $\boldsymbol{R}$ of CIPS can be written as:

$$R(A') = \begin{bmatrix} a & 0 & c \\ 0 & b & 0 \\ c & 0 & d \end{bmatrix},\ R(A'') = \begin{bmatrix} 0 & e & 0 \\ e & 0 & f \\ 0 & f & 0 \end{bmatrix}.$$

Assume the polarization vectors of incident light $\overrightarrow{e_i} = (x_i \quad y_i \quad z_i)$ and scattered light $\overrightarrow{e_s} = (x_s \quad y_s \quad z_s)$, the intensity of Raman modes can be expressed by:

$$I(A') = \left|\overrightarrow{e_i}A'\overrightarrow{e_s}^T\right|^2 = \left|(x_i \quad y_i \quad z_i)\begin{pmatrix} a & 0 & c \\ 0 & b & 0 \\ c & 0 & d \end{pmatrix}\begin{pmatrix} x_s \\ y_s \\ z_s \end{pmatrix}\right|^2$$

$$= |a \cdot x_i x_s + b \cdot y_i y_s + c \cdot (z_i x_s + x_i z_s) + d \cdot z_i z_s|^2,$$

$$I(A'') = \left|\overrightarrow{e_i}A''\overrightarrow{e_s}^T\right|^2 = \left|(x_i \quad y_i \quad z_i)\begin{pmatrix} 0 & e & 0 \\ e & 0 & f \\ 0 & f & 0 \end{pmatrix}\begin{pmatrix} x_s \\ y_s \\ z_s \end{pmatrix}\right|^2$$

$$= |e \cdot (y_i x_s + x_i y_s) + f \cdot (z_i y_s + y_i z_s)|^2.$$

According to the conditions of experimental measurement, the polarization vectors of incident light and scattered light can be decided as:

$$\overrightarrow{e_i} = \left(\frac{1}{\sqrt{2}} \quad \pm\frac{1}{\sqrt{2}}i \quad 0\right),\ \overrightarrow{e_s} = (\cos\beta \quad \sin\beta \quad 0),$$

from which the Raman modes' intensities can be calculated by:

$$I(A') = \int_0^{\pi}\left|\overrightarrow{e_i}A'\overrightarrow{e_s}^T\right|^2 d\beta = \frac{1}{4}\pi(a^2 + b^2),$$

$$I(A'') = \int_0^{\pi}\left|\overrightarrow{e_i}A''\overrightarrow{e_s}^T\right|^2 d\beta = \frac{1}{2}\pi e^2.$$

**Supplementary Text 2**

High-pressure single-crystal diffraction data were collected at 0.15 GPa, 0.43 GPa, 1.59 GPa, 2.46 GPa, 2.97 GPa and 3.77 GPa using an optimised data collection strategy based on a high-pressure pre-experiment in CrysAlis Pro.(2) On increasing pressure to 3.77 GPa, the crystal changed colour from pale yellow to orange, indicative of a phase transition.

Data were integrated and reduced in CrysAlis Pro,(2) and an absorption correction was applied using SADABS.(3) The crystal structure at 0.15 GPa was solved using the atomic positions of the ambient pressure structure as starting coordinates. All other high-pressure structures up to 2.97 GPa were solved using the starting coordinates of the previous pressure point. The trigonal structure at 3.77 GPa was solved using the starting coordinates of the previously published structure of $AgInP_2S_6$ (CSD deposition number: 1706772).(4)

All structures were refined against $|F^2|$ using (5) in Olex2(6) and treated as an inversion twin, except for the centrosymmetric trigonal structure. 1,2 and 1,3-distances of the $P_2S_6$ unit were restrained to those values measured under ambient conditions, while all metal bonds and angles were refined freely. Up to 2.97 GPa, the $Cu^I$ ion was modelled over two positions with free occupancies summing to one. The low completeness of the diffraction data prevented refinement of anisotropic displacement parameters, so all atoms were refined isotropically. At 3.77 GPa, the $Cu^I$ was refined over three positions (two of which were symmetry independent), with freely refined occupancy summing to one.

The compression experiment was repeated on a second crystal of $CuInP_2S_6$, which was characterized at ambient pressure, 1.10 GPa, 1.89 GPa, 2.17 GPa, 3.35 GPa and 5.83 GPa using a standardised collection strategy that employs an $\omega$-scan in 17 positions of $\varphi$ and 13 positions of $\kappa$. The phase transition and piezochromism was observed at 5.83 GPa. All data processing and structural refinement were performed as described previously. Only the unit cell dimensions could be determined at 5.83 GPa due to the low resolution of the diffraction data (~1.5 Å).

**Table S1.** Raman active phonon modes and their corresponding scattering intensity based on calculated Raman tensors at 0 K. Corresponding experimental frequencies are given in parentheses. The major Raman peaks discussed in the main text are highlighted in red.

| **No.** | **Raman Shift ($cm^{-1}$)** | **Raman tensor** | | | | | | **Mode** | **Intensity depending on polarization vectors of (incident, scattered) light** | | | |
|---|---|---|---|---|---|---|---|---|---|---|---|---|
| | | **a** | **b** | **c** | **d** | **e** | **f** | | **(100, 100)** | **(010, 010)** | **(001, 001)** | **(110,110)** |
| **60** | 571.340 | -- | -- | -- | -- | 0 | 0 | **A''** | -- | -- | -- | 0 |
| **59** | 567.190 | 2.39788 | -3.01767 | -1.40768 | 0.64610 | -- | -- | **A'** | 5.74982 | 9.10631 | 0.41745 | 0.38414 |
| **58** | 560.542 | -0.01286 | -1.80386 | -0.39907 | 0.76715 | -- | -- | **A'** | 1.65E-04 | 3.25391 | 0.58852 | 3.30048 |
| **57** | 560.534 | -- | -- | -- | -- | -2.82334 | -1.59542 | **A''** | -- | -- | -- | 10.18147 |
| **56** | 529.104 | 11.99121 | 13.16349 | -0.09671 | 2.74078 | -- | -- | **A'** | 143.78922 | 173.27749 | 7.51189 | 632.75919 |
| **55** | 528.588 | -- | -- | -- | -- | -0.07182 | 0.16210 | **A''** | -- | -- | -- | 0.1051 |
| **54** | 523.992 | -- | -- | -- | -- | -0.75910 | 1.85711 | **A''** | -- | -- | -- | 13.79546 |
| **53** | 523.629 | -5.61103 | 0.25967 | -2.12797 | -0.23920 | -- | -- | **A'** | 31.48365 | 0.06743 | 0.05722 | 28.63702 |
| **52** | 522.273 | -- | -- | -- | -- | 2.54793 | -1.83236 | **A''** | -- | -- | -- | 13.4301 |
| **51** | 521.981 | -4.70597 | -4.68141 | 0.97746 | -1.22373 | -- | -- | **A'** | 22.14619 | 21.91556 | 1.49751 | 88.12289 |
| **50** | 438.692 | -- | -- | -- | -- | -0.31518 | -0.45987 | **A''** | -- | -- | -- | 0.84594 |
| **49** | 430.851 | -0.01510 | -0.02149 | -0.00436 | -0.02672 | -- | -- | **A'** | 2.28E-04 | 4.62E-04 | 7.14E-04 | 0.00134 |
| **48** | 358.642 | -- | -- | -- | -- | -0.08730 | -0.11573 | **A''** | -- | -- | -- | 0.05357 |
| **47** | 358.218 (375) | -14.64970 | -15.08450 | 0.05108 | -5.03126 | -- | -- | **A'** | 214.61361 | 227.54205 | 25.31359 | 884.12226 |
| **46** | 323.095 | -- | -- | -- | -- | -0.16434 | 1.79137 | **A''** | -- | -- | -- | 12.83604 |
| **45** | 322.777 (319) | -10.57418 | 9.73858 | 0.86830 | -0.72090 | -- | -- | **A'** | 111.81337 | 94.8399 | 0.5197 | 0.69824 |
| **44** | 322.368 | -- | -- | -- | -- | -10.79708 | -0.93824 | **A''** | -- | -- | -- | 3.52117 |
| **43** | 322.118 | 0.80064 | -0.64018 | -1.65190 | 0.12394 | -- | -- | **A'** | 0.64102 | 0.40983 | 0.01536 | 0.02575 |
| **42** | 309.384 | -- | -- | -- | -- | -0.60598 | 0.13588 | **A''** | -- | -- | -- | 0.07386 |
| **41** | 299.316 | -10.00409 | -10.20966 | -0.26552 | -4.96447 | -- | -- | **A'** | 100.0819 | 104.23711 | 24.64597 | 408.59577 |
| **40** | 284.553 | -- | -- | -- | -- | 3.26500 | 1.35652 | **A''** | -- | -- | -- | 7.36057 |
| **39** | 283.937 | -4.63435 | 2.12334 | -0.50146 | -1.16753 | -- | -- | **A'** | 21.47721 | 4.50859 | 1.36312 | 6.30516 |
| **38** | 278.154 | 4.32621 | -3.30818 | -1.98917 | 0.12050 | -- | -- | **A'** | 18.71613 | 10.94406 | 0.01452 | 1.03639 |
| **37** | 278.113 | -- | -- | -- | -- | 3.26262 | -1.26840 | **A''** | -- | -- | -- | 6.43538 |
| **36** | 271.369 (266) | -3.38593 | -4.04844 | 0.03786 | -5.71702 | -- | -- | **A'** | 11.46454 | 16.38983 | 32.68437 | 55.26984 |
| **35** | 269.936 | -- | -- | -- | -- | -1.29428 | 0.46529 | **A''** | -- | -- | -- | 0.86596 |
| **34** | 242.325 (239) | 9.19039 | 9.52370 | -0.19327 | 17.60768 | -- | -- | **A'** | 84.46325 | 90.70088 | 310.03035 | 350.21716 |
| **33** | 232.218 | -- | -- | -- | -- | 0.05386 | 0.37907 | **A''** | -- | -- | -- | 0.57478 |
| **32** | 212.919 | -- | -- | -- | -- | 1.34771 | 0.27484 | **A''** | -- | -- | -- | 0.30214 |
| **31** | 212.875 | 1.01629 | -2.47448 | 3.42618 | -1.19482 | -- | -- | **A'** | 1.03285 | 6.12303 | 1.4276 | 2.1263 |
| **30** | 211.553 (216) | -- | -- | -- | -- | 2.65688 | -3.62334 | **A''** | -- | -- | -- | 52.51439 |
| **29** | 211.224 | -2.48985 | 1.55219 | -1.91257 | -0.83366 | -- | -- | **A'** | 6.19935 | 2.40931 | 0.69498 | 0.8792 |
| **28** | 196.899 | -0.53610 | -1.30592 | -1.31312 | -0.33285 | -- | -- | **A'** | 0.28741 | 1.70542 | 0.11079 | 3.39304 |
| **27** | 195.063 | -- | -- | -- | -- | -0.24567 | 0.44221 | **A''** | -- | -- | -- | 0.78219 |
| **26** | 191.264 | 0.28121 | -0.55031 | 0.64276 | 0.73796 | -- | -- | **A'** | 0.07908 | 0.30284 | 0.54459 | 0.07241 |
| **25** | 189.777 | -- | -- | -- | -- | -0.48006 | -1.20858 | **A''** | -- | -- | -- | 5.84265 |
| **24** | 161.534 (164) | -5.04150 | -5.08557 | -0.15256 | 3.84499 | -- | -- | **A'** | 25.41671 | 25.86298 | 14.78397 | 102.55743 |
| **23** | 160.215 | -- | -- | -- | -- | 0.50267 | -0.28371 | **A''** | -- | -- | -- | 0.32196 |
| **22** | 153.672 | -1.46915 | -0.10334 | -0.38683 | 0.15747 | -- | -- | **A'** | 2.1584 | 0.01068 | 0.0248 | 2.47272 |
| **21** | 153.217 | -- | -- | -- | -- | -0.77578 | 0.52772 | **A''** | -- | -- | -- | 1.11393 |
| **20** | 150.912 | -0.32036 | 1.85010 | -0.97472 | -1.15427 | -- | -- | **A'** | 0.10263 | 3.42288 | 1.33234 | 2.34012 |
| **19** | 150.399 | -- | -- | -- | -- | -0.81032 | 0.25619 | **A''** | -- | -- | -- | 0.26254 |
| **18** | 117.078 | -- | -- | -- | -- | 1.38184 | 0.76088 | **A''** | -- | -- | -- | 2.31578 |
| **17** | 115.073 | -1.24582 | -0.26249 | -0.32194 | 0.19686 | -- | -- | **A'** | 1.55206 | 0.0689 | 0.03876 | 2.27498 |
| **16** | 113.215 | -- | -- | -- | -- | 0.47457 | -0.22062 | **A''** | -- | -- | -- | 0.19469 |
| **15** | 112.385 | 1.73721 | -1.02618 | -0.98458 | -0.93173 | -- | -- | **A'** | 3.0179 | 1.05304 | 0.86812 | 0.50557 |
| **14** | 109.834 | -- | -- | -- | -- | -0.37350 | -0.12758 | **A''** | -- | -- | -- | 0.0651 |
| **13** | 101.200 (103) | 5.21062 | 5.34334 | -0.08302 | -3.87209 | -- | -- | **A'** | 27.15053 | 28.55132 | 14.99306 | 111.38607 |
| **12** | 73.689 | -- | -- | -- | -- | -0.63024 | -0.13712 | **A''** | -- | -- | -- | 0.07521 |
| **11** | 73.279 (~73) | 0.96116 | -1.57648 | -0.21488 | 0.64617 | -- | -- | **A'** | 0.92383 | 2.4853 | 0.41753 | 0.37862 |
| **10** | 70.708 | -- | -- | -- | -- | 0.91173 | 0.09433 | **A''** | -- | -- | -- | 0.03559 |
| **9** | 69.287 | -- | -- | -- | -- | 0.61921 | -0.00151 | **A''** | -- | -- | -- | 9.11E-06 |
| **8** | 69.062 | -0.59211 | -0.26032 | -0.03369 | 0.27053 | -- | -- | **A'** | 0.3506 | 0.06777 | 0.07319 | 0.72665 |
| **7** | 47.356 (~35) | -0.24083 | 0.07663 | -0.69344 | -5.08399 | -- | -- | **A'** | 0.058 | 0.00587 | 25.84696 | 0.02696 |
| **6** | 42.524 | -- | -- | -- | -- | 0.00410 | -0.12907 | **A''** | -- | -- | -- | 0.06664 |
| **5** | 28.580 | -- | -- | -- | -- | 0.09678 | -0.02426 | **A''** | -- | -- | -- | 0.00235 |
| **4** | 28.123 | -0.02592 | -0.08413 | 0.01247 | 0.09000 | -- | -- | **A'** | 6.72E-04 | 0.00708 | 0.0081 | 0.01211 |

**Table S2a.** Abridged crystallographic data for $CuInP_2S_6$ ($M_r$ = 432.66 g mol$^{-1}$) during hydrostatic compression to 3.77 GPa in a pressure-transmitting medium of MeOH/EtOH (4:1 vol%).

| **Sample Code** | **CIPS-Ambient** | **CIPS-0.15GPa** | **CIPS-0.43GPa** | **CIPS-1.59GPa** |
|---|---|---|---|---|
| Pressure (GPa) | 0.00010135 | 0.15 | 0.43 | 1.59 |
| Colour | Pale-yellow | Pale-yellow | Pale-yellow | Pale-yellow |
| Crystal system, space group | Monoclinic, *Cc* | Monoclinic, *Cc* | Monoclinic, *Cc* | Monoclinic, *Cc* |
| *a*, *b*, *c* (Å) | 6.10132 (11), 10.5742 (2), 13.2021 (3) | 6.0871 (4), 10.5478 (4), 13.169 (9) | 6.0860 (12), 10.546 (2), 12.984 (3) | 6.0340 (12), 10.459 (2), 12.786 (3) |
| α, β, γ (°) | 90, 99.1171 (17), 90 | 90, 99.24 (2), 90 | 90, 99.29 (3), 90 | 90, 99.48 (3), 90 |
| $V$ (Å$^3$) | 840.99 (3) | 834.5 (6) | 822.4 (3) | 795.9 (3) |
| $Z$ | 4 | 4 | 4 | 4 |
| μ (mm$^{-1}$) | 7.05 | 7.10 | 7.21 | 7.45 |
| Crystal size (mm) | 0.073 × 0.140 × 0.195 | 0.073 × 0.140 × 0.195 | 0.073 × 0.140 × 0.195 | 0.073 × 0.140 × 0.195 |
| Absorption correction | Multi-scan *SADABS2016*/2 | Multi-scan *SADABS2016*/2 | Multi-scan *SADABS2016*/2 | Multi-scan *SADABS2016*/2 |
| $T_{min}$, $T_{max}$ | 0.639, 0.746 | 0.507, 0.745 | 0.468, 0.730 | 0.403, 0.745 |
| No. of measured, independent and observed [$I > 2\sigma(I)$] reflections | 15553, 2780, 2459 | 5312, 616, 524 | 3248, 544, 427 | 6072, 465, 387 |
| $R_{int}$ | 0.037 | 0.068 | 0.127 | 0.152 |
| $\theta_{max}$ (°) | 32.1 | 26.4 | 25.0 | 23.2 |
| $(\sin \theta/\lambda)_{max}$ (Å$^{-1}$) | 0.748 | 0.626 | 0.595 | 0.555 |
| $R[F^2 > 2\sigma(F^2)]$, $wR(F^2)$, $S$ | 0.026, 0.056, 1.13 | 0.058, 0.193, 1.25 | 0.074, 0.263, 1.07 | 0.092, 0.287, 1.23 |
| No. of reflections | 2780 | 616 | 544 | 465 |
| No. of parameters | 100 | 45 | 45 | 45 |
| No. of restraints | 2 | 22 | 22 | 22 |
| $\Delta\rho_{max}$, $\Delta\rho_{min}$ (e Å$^{-3}$) | 0.84, −0.47 | 2.02, −1.47 | 2.16, −1.85 | 3.70, −2.50 |

**Table S2b.** Abridged crystallographic data for $CuInP_2S_6$ ($M_r$ = 432.66 g mol$^{-1}$) during hydrostatic compression to 3.77 GPa in a pressure-transmitting medium of MeOH/EtOH (4:1 vol%).

| | **CIPS-2.46GPa** | **CIPS-2.97GPa** | **CIPS-3.77GPa** |
|---|---|---|---|
| Pressure (GPa) | 2.46 | 2.97 | 3.77 |
| Colour | Pale-yellow | Pale-yellow | Orange |
| Crystal system, space group | Monoclinic, *Cc* | Monoclinic, *Cc* | Trigonal, $P\bar{3}1c$ |
| *a*, *b*, *c* (Å) | 6.0050 (12), 10.418 (2), 12.658 (3) | 5.9980 (12), 10.406 (2), 12.598 (3) | 5.9830 (4), 5.9830 (4), 12.094 (13) |
| α, β, γ (°) | 90, 99.58 (3), 90 | 90, 99.63 (3), 90 | 90, 90, 120 |
| *V* (Å$^3$) | 780.8 (3) | 775.2 (3) | 374.9 (4) |
| *Z* | 4 | 4 | 2 |
| μ (mm$^{-1}$) | 7.59 | 7.65 | 7.91 |
| Crystal size (mm) | 0.073 × 0.140 × 0.195 | 0.073 × 0.140 × 0.195 | 0.073 × 0.140 × 0.195 |
| Absorption correction | Multi-scan *SADABS2016*/2 | Multi-scan *SADABS2016*/2 | Multi-scan *SADABS2016*/2 |
| $T_{min}$, $T_{max}$ | 0.549, 0.733 | 0.603, 0.745 | 0.617, 0.745 |
| No. of measured, independent and observed [$I > 2\sigma(I)$] reflections | 2874, 445, 398 | 2920, 439, 395 | 2875, 76, 73 |
| $R_{int}$ | 0.080 | 0.047 | 0.034 |
| $\theta_{max}$ (°) | 23.3 | 23.3 | 23.1 |
| $(\sin \theta/\lambda)_{max}$ (Å$^{-1}$) | 0.557 | 0.557 | 0.551 |
| $R[F^2 > 2\sigma(F^2)]$, $wR(F^2)$, $S$ | 0.114, 0.339, 1.60 | 0.107, 0.315, 1.53 | 0.098, 0.200, 1.24 |
| No. of reflections | 445 | 439 | 76 |
| No. of parameters | 37 | 37 | 11 |
| No. of restraints | 22 | 22 | 5 |
| $\Delta\rho_{max}$, $\Delta\rho_{min}$ (e Å$^{-3}$) | 5.63, −2.24 | 5.85, −2.80 | 1.00, −0.75 |

**Table S3a.** Abridged crystallographic data for $CuInP_2S_6$ ($M_r$ = 432.66 g mol$^{-1}$) during hydrostatic compression to 5.83 GPa in a pressure-transmitting medium of MeOH/EtOH (4:1 vol%). Continued overleaf…

| **Sample Code** | **CIPS-2-Ambient** | **CIPS-2-1.10GPa** | **CIPS-2-1.89GPa** |
|---|---|---|---|
| Pressure (GPa) | 0.00010135 | 1.10 | 1.89 |
| Colour | Pale-yellow | Pale-yellow | Pale-yellow |
| Crystal system, space group | Monoclinic, *Cc* | Monoclinic, *Cc* | Monoclinic, *Cc* |
| $a$, $b$, $c$ (Å) | 6.0962 (1), 10.5704 (2), 13.1906 (2) | 6.0719 (5), 10.5120 (4), 12.95 (1) | 6.0446 (5), 10.4598 (4), 12.787 (9) |
| α, β, γ (°) | 99.120 (2) | 99.43 (3) | 99.44 (3) |
| $V$ (Å$^3$) | 839.25 (3) | 815.4 (6) | 797.5 (6) |
| $Z$ | 4 | 4 | 4 |
| μ (mm$^{-1}$) | 7.03 | 7.21 | 7.42 |
| Crystal size (mm) | 0.43 × 0.34 × 0.07 | 0.43 × 0.34 × 0.07 | 0.43 × 0.34 × 0.07 |
| Absorption correction | Multi-scan *SADABS2016*/2 | Multi-scan *SADABS2016*/2 | Multi-scan *SADABS2016*/2 |
| $T_{min}$, $T_{max}$ | 0.509, 0.746 | 0.494, 0.745 | 0.629, 0.737 |
| No. of measured, independent and observed [$I > 2\sigma(I)$] reflections | 10346, 2728, 2527 | 3550, 586, 490 | 3707, 574, 487 |
| $R_{int}$ | 0.038 | 0.076 | 0.059 |
| $\theta_{max}$ (°) | 32.195 | 26.319 | 28.938 |
| $(\sin \theta/\lambda)_{max}$ (Å$^{-1}$) | 0.751 | 0.624 | 0.623 |
| $R[F^2 > 2\sigma(F^2)]$, $wR(F^2)$, $S$ | 0.022, 0.060, 0.89 | 0.096, 0.332, 1.52 | 0.099, 0.300, 1.37 |
| No. of reflections | 2728 | 586 | 574 |
| No. of parameters | 101 | 45 | 45 |
| No. of restraints | 2 | 22 | 22 |
| $\Delta\rho_{max}$, $\Delta\rho_{min}$ (e Å$^{-3}$) | 0.65, −1.22 | 5.19, −3.43 | 5.62, −2.68 |

**Table S3b.** Abridged crystallographic data for $CuInP_2S_6$ ($M_r$ = 432.66 g mol$^{-1}$) during hydrostatic compression to 5.83 GPa in a pressure-transmitting medium of MeOH/EtOH (4:1 vol%).

| **Sample Code** | **CIPS-2-2.17GPa** | **CIPS-2-3.35GPa** | **CIPS-2-5.83GPa** |
|---|---|---|---|
| Pressure (GPa) | 2.17 | 3.35 | 5.83 |
| Colour | Pale-yellow | Pale-yellow | Orange |
| Crystal system, space group | Monoclinic, *Cc* | Monoclinic, *Cc* | Trigonal, $P\bar{3}1c$ |
| *a*, *b*, *c* (Å) | 6.0290 (6), 10.4265 (6), 12.733 (11) | 6.0034 (9), 10.3584 (7), 12.544 (15) | 5.9293(10), 5.9293(10), 11.83(2) |
| α, β, γ (°) | 99.40 (3) | 99.35 (5) | 90, 90, 120 |
| *V* (Å$^3$) | 789.6 (7) | 769.7 (10) | 360.3(7) |
| *Z* | 4 | 4 | 2 |
| μ (mm$^{-1}$) | 7.57 | 7.70 | 8.23 |
| Crystal size (mm) | 0.43 × 0.34 × 0.07 | 0.43 × 0.34 × 0.07 | 0.43 × 0.34 × 0.07 |
| Absorption correction | Multi-scan *SADABS2016*/2 | Multi-scan *SADABS2016*/2 | Multi-scan *SADABS2016*/2 |
| $T_{min}$, $T_{max}$ | 0.555, 0.745 | 0.475, 0.745 | 0.477, 1.000 |
| No. of measured, independent and observed [$I > 2\sigma(I)$] reflections | 3761, 571, 487 | 3764, 562, 466 | 2376, 61, 57 |
| $R_{int}$ | 0.057 | 0.071 | 0.064 |
| $\theta_{max}$ (°) | 26.325 | 25.999 | 20.581 |
| $(\sin \theta/\lambda)_{max}$ (Å$^{-1}$) | 0.625 | 0.621 | n/a |
| $R[F^2 > 2\sigma(F^2)]$, $wR(F^2)$, $S$ | 0.086, 0.255, 1.18 | 0.105, 0.296, 1.27 | n/a |
| No. of reflections | 571 | 562 | n/a |
| No. of parameters | 45 | 37 | n/a |
| No. of restraints | 22 | 22 | n/a |
| $\Delta\rho_{max}$, $\Delta\rho_{min}$ (e Å$^{-3}$) | 4.78, −3.02 | 5.59, −3.33 | n/a |

Reference

1. Y. Peter, M. Cardona, *Fundamentals of Semiconductors*. Graduate Texts in Physics (Springer Science & Business Media, 2010).
2. R. O. Diffraction. (Rigaku Corporation Wroclaw, Poland, 2019).
3. G. Sheldrick, SADABS Bruker Axs Inc. *Madison, Wisconsin, USA*, (2007).
4. Z. Ouili, A. Leblanc, P. Colombet, Crystal structure of a new lamellar compound: Ag12In12PS3. *J Solid State Chem.* **66**, 86-94 (1987).
5. G. M. Sheldrick, Crystal structure refinement with SHELXL. *Acta Cryst C* **71**, 3-8 (2015).
6. O. V. Dolomanov, L. J. Bourhis, R. J. Gildea, J. A. K. Howard, H. Puschmann, OLEX2: a complete structure solution, refinement and analysis program. *J Appl Cryst* **42**, 339-341 (2009).